\documentclass[aps,pra,twocolumn,groupedaddress,floatfix]{revtex4-2}

\usepackage{graphicx}
\usepackage{amsmath}
\usepackage{amsfonts}
\usepackage{bm}
\usepackage{xfrac}
\usepackage[utf8]{inputenc}

\DeclareMathOperator{\grad}{grad}
\DeclareMathOperator{\curl}{curl}      
\newcommand{\mr}[1]{\ensuremath{\mathrm{#1}}}

\begin{document}

\title{Bayesian Estimation of Experimental Parameters in Stochastic Inertial Systems: Theory, Simulations, and Experiments with Objects Levitated in Vacuum 
}
\author{M. \v{S}iler}
\email{siler@isibrno.cz}
\author{V. Svak}
\author{A. Jon\'a\v{s}}
\author{S. H. Simpson}
\author{O. Brzobohat\'{y}}
\author{P. Zem\'{a}nek}
\affiliation{The Czech Academy of Sciences, Institute of Scientific Instruments, Kr\'{a}lovopolsk\'{a} 147, 612 64 Brno, Czech Republic}

\begin{abstract}
High-quality nanomechanical oscillators can sensitively probe force, mass, or displacement in experiments bridging the gap between the classical and quantum domain. 
Dynamics of these stochastic systems is inherently determined by the interplay between acting external forces, viscous dissipation, and random driving by the thermal environment. 
The importance of inertia then dictates that both position and momentum must, in principle, be known to fully describe the system, which makes its quantitative experimental characterization rather challenging. 
We introduce a general method of Bayesian inference of the force field and environmental parameters in stochastic inertial systems that operates solely on the time series of recorded noisy positions of the system. 
The method is first validated on simulated trajectories of model stochastic harmonic and anharmonic oscillators with damping. 
Subsequently, the method is applied to experimental trajectories of particles levitating in tailored optical fields and used to characterize the dynamics of particle motion in a nonlinear Duffing potential, a static or time-dependent double-well potential, and a  non-conservative force field. 
The presented inference procedure does not make any simplifying assumptions about the nature or symmetry of the acting force field and provides robust results with trajectories two orders of magnitude shorter than those typically required by alternative inference schemes. 
In addition to being a powerful tool for quantitative data analysis, it can also guide experimentalists in choosing appropriate sampling frequency (at least 20 measured points per single characteristic period) and length of the measured trajectories (at least 10 periods) to estimate the force field  and environmental characteristics with a desired accuracy and precision.
\end{abstract}

\maketitle

\section{Introduction}

Temporal evolution of the state of a microscopic dynamical system interacting with a heat reservoir results from the combined action of deterministic forces and random thermal noise~\cite{ChandrasekharRMP43, Wang_RMP_1945}. On the microscale, inertial effects are typically negligible in comparison with dissipative viscous forces and the stochastic system of interest can be described by {\em overdamped} first-order Langevin dynamics. 
Recently, various types of high-quality nanomechanical oscillators have been systematically explored for applications in ultrasensitive detection of force~\cite{MoserNatNanotech13, RanjitPRA16}, mass~ \cite{LassagneNanoLett08}, or displacement~\cite{ TeufelNatNanotech09} and for experimental testing of the laws of quantum mechanics in previously inaccessible parameter regimes~\cite{TeufelNature2011,Tebbenjohanns_PRL_2020}. For such nanomechanical oscillators suspended in vacuum, where the dissipative interaction with the environment is strongly reduced, inertia becomes essential. Consequently, proper description of these systems involves second-order dynamics, formally captured by the {\em underdamped} Langevin equation. 

Micro- and nanoparticles optically levitated in vacuum represent a unique experimental platform that offers a large degree of control over the characteristic parameters of the force field and thermal environment that govern the stochastic dynamics of the system~\cite{Millen_RPP_2020}. Hence, they are ideally suited for systematic quantitative studies of stochastic phenomena in the presence of both strong and weak dissipation. 
Specifically, viscous damping can be directly varied over many orders of magnitude by adjusting the ambient pressure in the vacuum chamber from the standard atmospheric pressure down to $10^{-10}$\,Pa. Effective kinetic temperature of the particle's center-of-mass motion can be reduced well into the sub-Kelvin range by various methods of autonomous~\cite{DelicScience20} or feedback-assisted~\cite{Gieseler2012Subkelvin,TebbenjohannsPRL19} cooling. 
The inertial mass of a levitated particle is proportional to its volume; since the particle size typically ranges from tens of nanometers to micrometers, available masses span about six orders of magnitude. Furthermore, spatial and temporal dependence of force fields acting upon the particle can be dynamically tailored by shaping the intensity and/or phase profile of laser beams used for optical levitation~\cite{CurtisOPTCOMM02,CizmarLPL11,SilerPRL18}. 
\begin{figure*}[thb]
  \includegraphics[width=0.8\textwidth]{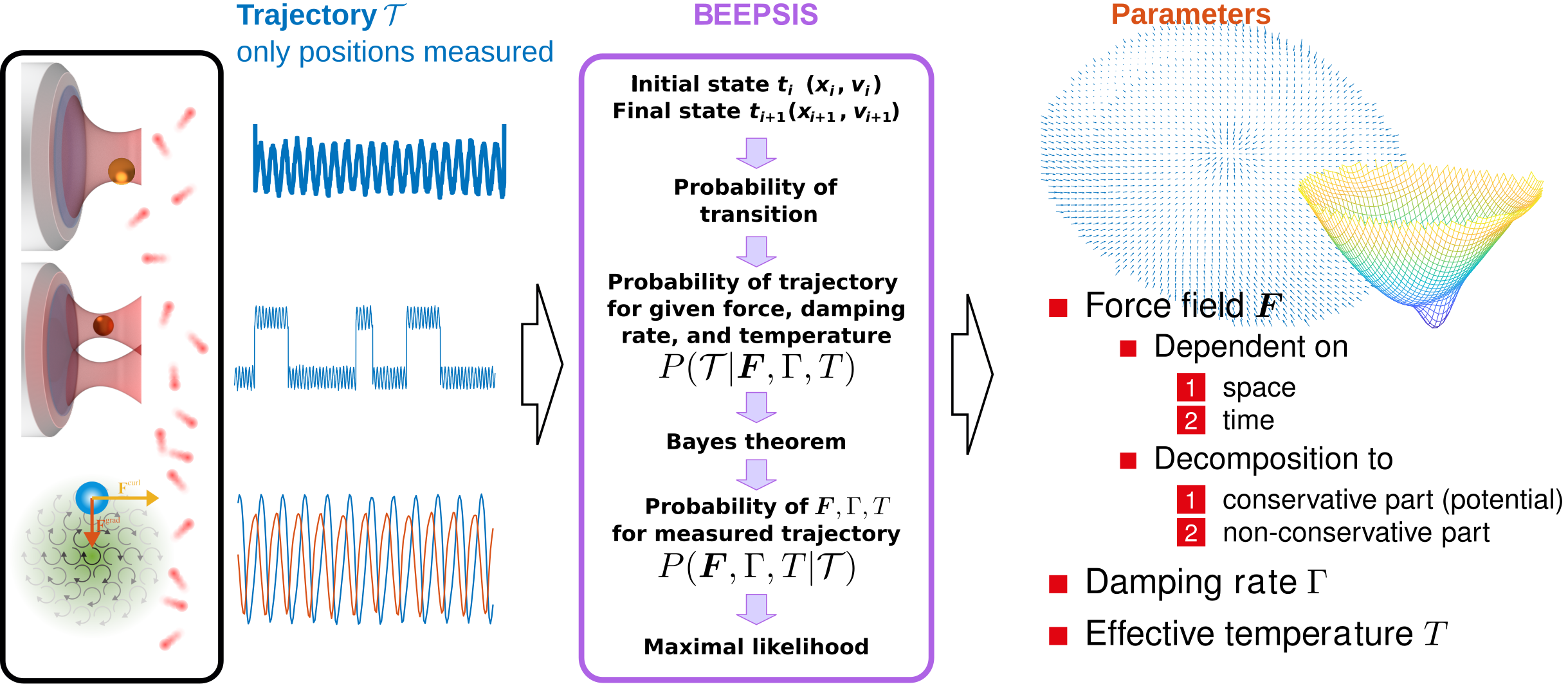}
   \caption{Schematic illustration of the overall operating workflow of BEEPSIS. Experimental trajectories $\mathcal T$ are used to express the probability distribution $P\!\left( \bm{F},\Gamma,T  | \mathcal T \right)$ of the characteristic parameters of the system (force field $\bm{F}$, viscous damping rate $\Gamma$, and thermodynamic temperature $T$) for the actual measured set of discrete positions of the system (see text for details).
   }
   \label{fig:principle}
 \end{figure*} 

Experimental trajectories recorded with sufficient temporal and spatial resolution serve as the starting point for inferring the parameters of the system and its surroundings – in particular, the effective damping coefficient, the effective temperature, and the characteristic parameters of the force field – that enter into the Langevin equation of motion. 
First-order stochastic dynamics of overdamped systems – also known as the Brownian motion – can be fully described by a series of positions measured at discrete times~\cite{SiegertPLA1998, RagwitzPRL2001}. 
There exist multiple methods for extracting information about optically manipulated particles and their environment from the trajectories recorded in the overdamped regime~\cite{Jonesbook15, GieselerAOP2021}. Spatial profile of a conservative force can be reconstructed using the Boltzmann distribution~\cite{FlorinAPA98} or - for a harmonic potential - the equipartition theorem. 
Analysis of the time-correlated properties of the motion of a Brownian particle in the harmonic potential via calculating the autocorrelation function or the power spectral density of the particle’s position provides access to both potential stiffness and viscous damping coefficient~\cite{BergSorensenRSI04}.
More recent approaches to quantitative characterization of overdamped stochastic systems include Bayesian inference~\cite{RichlyOE13}, maximum-likelihood estimation~\cite{PerezGarciaNatComm18}, and application of information-theory concepts~\cite{FrishmanPRX20}. These methods allow for determination of viscous damping coefficient along with the potential profiles and possibly even non-conservative components of the force field.
  
When the dynamics of the studied system is of second order, its full description requires knowledge of a pair of suitable conjugate variables, typically positions and velocities~\cite{Strogatz}. If these conjugate variables are both independently experimentally accessible, it is possible to correctly represent the evolution of the studied stochastic system by a discretized form of a memoryless continuous-time process, which is amenable to the standard Markov analysis. 
However, in a typical experimental scenario, only the system positions in discrete times are directly measured with a certain level of error and the instantaneous velocities must be estimated by numerically differentiating the recorded trajectories. This introduces memory into the discretized Langevin equation, making it inherently more complex to solve than its first-order counterpart. 
In addition, measurement errors are amplified by differentiation and generally lead to systematic biases in the parameter estimation~\cite{LehlePRE2015}. For these reasons, quantitative inference in stochastic inertial systems subject to general nonlinear force fields is rather challenging and reliable strategies for tackling this problem have been reported only recently~\cite{FerrettiPRX20, BrucknerPRL20}.

Here, we present a novel inference method for quantitative characterization of underdamped stochastic dynamics: Bayesian Estimation of Experimental Parameters in Stochastic Inertial Systems (BEEPSIS). 
BEEPSIS exploits discrete particle trajectories recorded in the presence of detection noise to find the local force acting on the particle at the experimentally observed positions and to determine the effective temperature and damping characteristics of the ambient environment. It can be applied to a general system with inertia subject to an external force field containing conservative, non-conservative, or time-varying components. 
The quantities of interest are numerically estimated using maximal-likelihood approach based on the Bayes theorem, assuming a known mass or normalizing the underlying Langevin equation to a unit mass. The overall workflow of BEEPSIS is schematically illustrated in Fig. \ref{fig:principle}. 
The likelihood function describing the studied system is expressed solely in terms of particle positions; it does not require discrete-time approximations for instantaneous velocities and/or accelerations. This approach is similar to the strategy adopted in~\cite{FerrettiPRX20}; however, our focus is especially on the processing of experimental data. Therefore, we analyze the requirements on the sampled trajectories necessary for high-fidelity inference of system  parameters and provide a data processing approach that is capable of dealing with big data sets. Furthermore, our formalism explicitly includes the effects of the position measurement error in the likelihood formula.
The article is organized as follows: In Section~\ref{sec:inference_theory}, we introduce our theoretical framework and derive the formula for the conditional probability of transition along the actual observed trajectory. We then show how this conditional probability can be used to infer the system parameters both in the absence and in the presence of measurement errors. 
In Section~\ref{sec:simul}, we study the robustness of BEEPSIS inference using simulated stochastic trajectories of harmonic oscillator and Duffing oscillator, which represent frequently used linear and nonlinear parametric models of optical traps. We analyze the accuracy and precision of the inference for a wide range of system parameters as functions of the number of periods included in the trajectory and the number of points sampled per period. This information is crucial to properly choose the sampling frequency and trajectory length to reliably determine the target parameters. 
Since our primary motivation is the analysis of data obtained from optical levitation experiments in vacuum, in Section~\ref{sec:experiments}, we apply BEEPSIS to the experimental trajectories of particles confined in optical force fields of increasing complexity. Specifically, we quantitatively characterize optical levitation in an anharmonic Duffing potential~\cite{FlajsmanovaSR20}, in a double-well, time-modulated potential, and in a two-dimensional non-conservative force field~\cite{SvakNC18}. Finally, in Section~\ref{sec:conclusion}, we summarize our findings and provide an outline for potential future developments of BEEPSIS.

\section{Bayesian inference in underdamped stochastic systems}
\label{sec:inference_theory}
\subsection{Probability of particle transition along an experimentally observed trajectory}
Let us consider a spherical particle of radius $a$ and mass $m$ that is immersed in a thermal bath with thermodynamic temperature $T$ and viscosity $\eta$ and moves in a general spatially dependent force field $\bm{F} (\bm{x})$, where $\bm{x}$ is the particle position vector. 
The motion of the particle can be described by the second-order Langevin equation (LE) \cite{Risken}
\begin{equation}
  m  \bm{\ddot x} + m \Gamma \bm{\dot x}  =  \bm{F}(\bm{x}) + \bm{\xi}(t),
  \label{eq:LE}
\end{equation} 
where $\Gamma$ is the viscous damping rate (Stokes drag coefficient $6 \pi \eta a$ over mass $m$) and $\xi$ is the random force originating in collisions with the molecules of the surrounding medium. 
$\xi$ is assumed to be the standard white noise with zero mean ($\langle \bm{\xi} \rangle = 0$) and time correlation given by
\begin{equation}
  \langle \bm{\xi}(t) \bm{\xi}(t')\rangle = 2k_BT m \Gamma \delta(t-t') \bm{\mathrm{I}},
  \label{eq:LEcov}
\end{equation} 
with $k_B$ being the Boltzmann constant and $\bm{\mathrm{I}}$ being the identity matrix. 
In principle, Eq.~(\ref{eq:LE}) can be applied both to an individual particle and to an ensemble of interacting particles~\cite{SvakOptica21,RieserScience22}. 
To handle such general scenarios, one can formally assume that the motion takes place in space with $N$ degrees of freedom and is characterized by an $N$-dimensional position vector $\bm{x}$. Furthermore, in the case of multiple interacting particles, mass $m$ and mobility $1/(m\Gamma)$ of a single particle are replaced by the appropriate mass and mobility tensors~\cite{DoiPolymerBook}.

Let us assume that the $N$-dimensional trajectories of the system are sampled at discrete, regular time steps of length $\tau=t_{i+1}-t_i$.  
The discretized formal solution of~(\ref{eq:LE}), represented by the position and velocity of the system  $(\bm{x}_{i+1}, \bm{v}_{i+1})$ at time $t_{i+1}$, can be expressed in terms of the phase-space coordinates $(\bm{x}_i, \bm{v}_i)$ at a previous time $t_i$ as~\cite{ErmakJCompPhys80}: 
\begin{widetext}
 \begin{eqnarray}
 \label{eq:fLEv}
 \bm{v}_{i+1} - \bm{v}_i\mr e^{-\Gamma \tau} - \frac{\bm F (\bm x_i)}{m\Gamma}\left(1- e^{-\Gamma \tau} \right) &=& \frac{1}{m}\int_0^\tau e^{-\Gamma (\tau-t')} \bm \xi (t_i+t') \mr d t',\\
 \label{eq:fLEx}
 \bm{x}_{i+1} -\bm x_i - \frac{\bm{v}_i}{\Gamma} \left(1-\mr e^{-\Gamma \tau}\right) - \frac{\bm F( \bm x_i)}{m\Gamma^2}\left(\Gamma \tau -1 + e^{-\Gamma \tau} \right) &=& \frac{1}{m\Gamma}\int_0^\tau \left[1-e^{-\Gamma (\tau-t')} \right]\bm \xi (t_i+t') \mr d t',
\end{eqnarray}
where we restricted the value of $\tau$ so that the force $\bm{F(x_i)}$ remains essentially constant in the time interval $(t_i, t_i+\tau)$. Following the procedure outlined in~\cite{GronbechJensenMolPhys13}, we can eliminate the velocity variable
and obtain the position update formula
\begin{eqnarray}
 \nonumber
 && \bm{x}_{i+1} -\bm{x}_i\left(1+ \mr e^{-\Gamma \tau}\right) + \bm{x}_{i-1}\mr e^{-\Gamma \tau} - \frac{\bm F(\bm x_i) \tau}{m\Gamma}\left(1-\mr e^{-\Gamma \tau}\right) = \\
 \label{eq:fLE2}
 &=& \frac{1}{m\Gamma} \left\{ \int_0^\tau \left[1-e^{-\Gamma (\tau-t')} \right]\bm \xi (t_i+t') \mr d t' + \mr e^{-\Gamma \tau} \int_0^\tau \left[e^{\Gamma t'} -1\right]\bm \xi (t_{i-1}+t') \mr d t'\right\},
\end{eqnarray}
\end{widetext}
where the assumption of constant $\bm{F(x_i)}$ was extended to the time interval $(t_i - \tau, t_i+\tau)$. Position update $\bm{x}_{i+1}$ calculated from~(\ref{eq:fLE2}) depends on two previous positions, $\bm{x}_{i}$ and $\bm{x}_{i-1}$, and on random force contributions (described by the stochastic integrals of $\bm \xi$) evaluated over two adjacent time intervals. Hence, $\bm{x}_{i+1}$ is not a Markovian quantity, as it displays memory of the system history.

Let us denote by  $P(\bm x_{L+1}, \bm {x}_L, \bm {x}_{L-1}, \dots, \bm{x}_2 | \bm{x}_1, \bm {x}_0,\bm{\phi})$ the conditional probability distribution of evolution of the particle position along the trajectory $\mathcal T = \{\bm {x}_0,\cdots,\bm x_{L+1}\}$, provided that the particle initially passed through positions $\bm {x}_0, \bm {x}_1$ under the set of parameters $\bm{\phi}=\{\bm{F},\Gamma,T,m\}$ that fully characterize the studied stochastic system ``particle + thermal bath''.
$P(\bm x_{L+1}, \bm {x}_L, \bm {x}_{L-1}, \dots, \bm{x}_2 | \bm{x}_1, \bm {x}_0,\bm{\phi}) \equiv P(\mathcal T| \bm{\phi})$ is a $L$-dimensional multivariate normal distribution for random variables $\bm x_{i+1}$ ($i\in \{1, \cdots, L\}$). The means of $\bm x_{i+1}$ are predicted from the previous positions $\bm x_i$, $\bm x_{i-1}$ and force $\bm F(\bm x_i)$ using~(\ref{eq:fLE2}), whereas the covariances of $\bm x_{i+1}$ describe correlations between the random forces in adjacent time intervals~\cite{SiviaBook,Risken,FerrettiPRX20}. 
$P(\mathcal T| \bm{\phi})$ can be expressed as  
\begin{equation}
 \begin{split}
  &P(\mathcal T | \bm{\phi}) = \frac{1}{\left[(2\pi)^{L} \mr{det} \mathbf C\right]^{N/2}}
  \\
  &\times\exp\left\{-\frac{1}{2} \sum_{n=x,y,\dots} \sum_{i,j=1}^{L}\bm{m}_{i,n} (\mathbf C^{-1})_{ij} \bm{m}_{j,n}\right\},
  \end{split}
  \label{eq:Pc}
\end{equation} 
where the first and second summation run over the $N$ degrees of freedom of the system and over the individual points of the system's trajectory with the total length of $(L+2)$ points, respectively. 
Furthermore, $\bm{m}$ is a $N$-dimensional vector of misfits or residuals~\cite{SiviaBook}
\begin{equation}
 \begin{split}
 \bm{m_i} \equiv & \,\, \bm{x}_{i+1} -\bm{x}_i\left(1+ \mr e^{-\Gamma \tau}\right) + \bm{x}_{i-1}\mr e^{-\Gamma \tau} 
 \\&- \frac{\bm{F(x_i)}\tau}{m\Gamma}\left(1-\mr e^{-\Gamma \tau}\right),
 \end{split}
 \label{eq:misfit}
\end{equation} 
and $\mathbf C$ is the covariance matrix of the random force. 
The misfits are equal to the deterministic left-hand side of Eq.~(\ref{eq:fLE2}) with $\bm{x}_{i+1}$ being the ``actual measured position'' and the remaining part being the ``predicted position''.
The covariance matrix with elements $\mathbf{C}_{ij} = \langle \bm{m}_i \bm{m}_j \rangle$ then describes correlations of misfits at  different times $\{t_i, t_j\}$ and can be evaluated using the right-hand side of Eq.~(\ref{eq:fLE2}).
It has the form of a tri-diagonal Toeplitz matrix ~\cite{FerrettiPRX20}
\begin{equation}
 \mathbf C = \left( 
    \begin{array}{cccccccc}
      a&b& 0& & \dots&0  \\
      b&a&b& & \\
      0&b&a&b& & \\
      \phantom{\ddots}&&\ddots&\ddots& \ddots& \\
      \vdots&&&b&a&b\\
      0&&&&b&a\\
    \end{array}
 \right), 
 \label{eq:Cmatrix}
\end{equation}
where the coefficients $a$ and $b$ are the covariances of noise terms that can be evaluated using It\^o isometry as (see Appendix~\ref{app:ito} for details)
\begin{eqnarray}
 a&=& 2 \frac{k_BT}{m \Gamma^2} \left[\Gamma \tau - 1 + \left( \Gamma \tau + 1 \right) \mr e^{-2\Gamma \tau} \right], \label{eq:Ca}\\
 b&=& \frac{ k_BT}{m \Gamma^2} \left[1 - 2\Gamma \tau \mr e^{-\Gamma \tau} - \mr e^{-2\Gamma \tau}\right].
 \label{eq:Cb}
\end{eqnarray}
For short time steps $\tau$ (i.e., $\Gamma \tau \ll 1$), one may perform the Taylor expansion of the exact exponential terms in Eq. (\ref{eq:misfit}), (\ref{eq:Ca}), and  (\ref{eq:Cb}) (see Appendix \ref{app:taylor}).
In this approximation, the dependence of $P(\mathcal T | \bm{\phi})$ on the instantaneous velocity and acceleration, estimated from the observed positions by the central difference scheme, has the form similar to that reported in~\cite{FerrettiPRX20}. 

In the cases where the evolution of the full phase-space trajectory is known, i.e., both positions $\bm{x}_{i}$ and instantaneous velocities $\bm{v}_{i}$ at discrete times $t_i$ are independently accessible, the system characterized by the coordinate pairs $(\bm{x}_{i},\bm{v}_{i})$ becomes Markovian~\cite{LehlePRE2015}. Consequently, simpler forms of the transition probability distribution~(\ref{eq:Pc}) containing only diagonal elements of the covariance matrix can be adopted (see  Appendix~\ref{app:velocpdf} and~\cite{Risken,ChandrasekharRMP43,FerrettiPRX20}).

\subsection{Estimation of parameters of the studied stochastic system from Bayes' theorem}
Transition probability distribution~(\ref{eq:Pc}) forms the foundation for estimating the parameters of the studied stochastic system. Applying Bayes' theorem, we can calculate the probability distribution $P\left( \bm{\phi}| \mathcal T \right)$ of the system parameters $\bm{\phi}$ conditioned on the evolution of the system along the actual observed trajectory $\mathcal T$ as~\cite{TurkcanBJ12,SiviaBook,FerrettiPRX20}
\begin{equation}
P\left( \bm{\phi}| \mathcal T \right) \propto  P\left(\mathcal T | \bm{\phi}\right)  P_0\left( \bm{\phi}\right).
\label{eq:Bayes}
\end{equation} 
Here, $P_0\left( \bm{\phi} \right)$ is the prior probability distribution of the parameters and $P\left(\mathcal T | \bm{\phi}\right)$ represents the likelihood quantifying the compatibility of the observed trajectory with a particular parameter set $\bm{\phi}$.
Upon choosing a uniform prior (either for simplicity or due to the lack of more detailed information)~\cite{TurkcanBJ12,FerrettiPRX20}, both sides of~(\ref{eq:Bayes}) differ only by a multiplicative constant. Consequently, the maximum of $P\left( \bm{\phi}| \mathcal T \right)$, which corresponds to the most probable set of system parameters $\bm{\phi}$ compatible with the observed trajectory $\mathcal T$, directly coincides with the maximum-likelihood estimator (MLE)~\cite{FerrettiPRX20}.
Following the standard MLE procedure, we can then define the negative log-likelihood $\mathcal L$ as
\begin{equation}
  \begin{split}
 \mathcal{L} =& -\log P\left(\mathcal T | \bm{\phi}\right) = \frac{LN}{2} \log 2\pi + \frac{N}{2} \log \det \bm{C} \\ &+ \frac12 \sum_{n=x,y,\dots} \sum_{i,j=1}^{L}\bm{m}_{i,n} (\mathbf C^{-1})_{ij} \bm{m}_{j,n},
 \end{split}
  \label{eq:bsum}
\end{equation} 
which can be numerically minimized to yield the desired system parameters $\bm{\phi}$.
The force field $\bm {F}(\bm{x_i},t_i)$ at discrete positions $\bm{x_i}$ and times $t_i$, which represents a subset of $\bm{\phi}$, can be described either by a parameterized analytical formula or by an appropriate interpolation scheme. 
Examples of the first approach include harmonic oscillator with stiffness $\kappa$ leading to force $F=-\kappa x$, nonlinear Duffing oscillator~\cite{GieselerNatPhys13,FlajsmanovaSR20}, double-well potential, or various time-dependent periodic and aperiodic force fields used in modeling directed Brownian transport~\cite{RevModPhysHanggi2009}.
Alternatively, we can describe a non-parametric force field by its values $\bm {F}(x_i)$ at the points $x_i$ of the observed trajectory of the system and use some form of interpolation to get the values of the force over a finer grid. 
The simplest example of a possible interpolation scheme is the ``binning'' method, in which the force is assumed to be constant over a bin of given constant spatial dimensions. 
More refined approaches include, for example, linear, spline, B\'ezier, or Lanczos interpolation schemes. 

In many practical situations, the trajectory length $L$ exceeds $10^6$ points; therefore, direct inversion of the covariance matrix $\mathbf{C}$ and calculation of the logarithm of its determinant, required to minimize Eq.~(\ref{eq:bsum}), may be impossible due to the computer memory and machine precision limitations. 
In such cases, the determinant of $\mathbf{C}$ can be calculated using the Laplace expansion and linear homogeneous recurrence relation as~\cite{ToeplitzDet}
\begin{equation}
 \mr{det} \mathbf C = \left\{ 
   \begin{array}{ll}
      \frac{1}{d} \left[ \left(\frac{a+d}{2} \right)^{L\!+\!1} -  \left(\frac{a-d}{2} \right)^{L\!+\!1} \right] & \mathrm{if}\ a^2\neq4b^2\\
      (L+1) \left(\frac{a}{2} \right)^{L} &\mathrm{if}\ a^2 = 4b^2,\\
   \end{array}
 \right.
  \label{eq:Cdet}
\end{equation} 
where $d = \sqrt{a^2-4b^2}$.
The ratio of the on- and off-diagonal elements $a,b$ of $\mathbf{C}$ is approximately $a/b \simeq 4$ (see Eq.~(\ref{eq:Ca}) and (\ref{eq:Cb}) and Appendix \ref{app:taylor}), which implies $d \simeq \sqrt{3}a/2$. 
Consequently, for long trajectories, the second term in~(\ref{eq:Cdet}), proportional to $(a-d)^{L+1}$, can be neglected and 
\begin{equation}
  \log (\det \mathbf{C}) \simeq -\log d + (L+1) \log\left(\frac{a + d}{2} \right). 
 \label{eq:logdetC}
\end{equation} 
The inversion of $\mathbf{C}$ can be carried out using either the eigenvalues of $\mathbf{C}$~\cite{FerrettiPRX20} or the Chebyshev polynomials of the second kind~\cite{DaFonsecaLinAlgApp01}. 
However, one may entirely avoid calculating $\mathbf{C}^{-1}$ by realizing that the sum involving $\mathbf{C}^{-1}$ in~(\ref{eq:bsum}) can be rewritten as 
\begin{equation}
 \sum_{i,j=1}^{L}\bm{m}_{i,n} (\mathbf C^{-1})_{ij} \bm{m}_{j,n} =  \sum_{i=1}^{L}\bm{m}_{i,n}  \bm{n}_{i,n},
 \label{eq:Csystem}
\end{equation} 
where $\bm{n}  = \mathbf C^{-1} \bm{m}$, i.e., $\bm{n}$ is a solution of the system of linear equations $\mathbf C \bm{n} = \bm{m}$. 
This sparse system can be easily numerically solved for $\bm{n}$.

\subsection{Effect of detection uncertainties}
\label{sec:error}
The procedure for the Bayesian inference of system's parameters introduced in the previous section assumes that the observed trajectory points $\bm x_i$ represent the true positions of the system.
However, this is typically not the case and the actual measured positions $\bm s_i = \bm x_i + \bm \psi_i$ contain also some experimental uncertainty $\bm \psi_i$, which is usually modeled as a Gaussian random variable with zero mean and variance $\sigma^2$, uncorrelated in time.
In order to calculate the transition probability distribution for the observed noisy trajectory $\mathcal T_s = \{\bm {s}_0,\cdots,\bm s_{L+1}\}$, we start with Eq.~(\ref{eq:fLE2}), replace $\bm{x}_i$  by $(\bm{s_i}-\bm \psi_i)$, and use the first-order Taylor expansion of $\bm F(\bm s_i - \bm \psi_i)$. 
By further employing the fact that the random Langevin force and detection noise are uncorrelated, we obtain the modified transition probability distribution of the trajectory $\mathcal T_s$ recorded in the presence of detection uncertainty 
\begin{widetext}
\begin{equation}
   P(\mathcal T_s | \bm{\phi} ) = \prod_{n=x,y,\dots} \frac{1}{\left[{(2\pi)^{L} \mr{det} \left(\mathbf C + \mathbf \Psi_n\right)}\right]^{1/2}} \exp\left\{-\frac{1}{2} \sum_{i,j=1}^{L}\bm{m'}_{i,n} [(\mathbf C + \mathbf \Psi_n)^{-1}]_{ij} \bm{m'}_{j,n}\right\},
  \label{eq:Pcerr}
\end{equation} 
where $\bm {m'}_i$ is the vector of misfits (\ref{eq:misfit}) obtained for the measured noisy positions $\{\bm s_{i-1}, \bm s_i, \bm s_{i+1}\}$, the covariance matrix $\mathbf{C}$ given by~(\ref{eq:Cmatrix}) -- (\ref{eq:Cb}) is unchanged, and $\mathbf \Psi_n$ is the covariance matrix of the detection noise along the $n$-th position coordinate that has a symmetric band structure with nonzero elements located on the main diagonal and the four adjacent diagonals. In particular, the elements of $\mathbf \Psi_n$ can be expressed as 
\begin{eqnarray}
 \Psi_{i,i,n} &=& 2 \sigma^2 \left[1+\mr e^{-\Gamma \tau} +\mr e^{-2 \Gamma \tau} + \frac{(1-\mr e^{-2 \Gamma \tau})\tau}{m\Gamma}\bm{j}_n(\bm{s}_i) + \frac12\left(\frac{(1-\mr e^{- \Gamma \tau})\tau}{m\Gamma}\bm{j}_n(\bm{s}_i)\right)^2 \right],\\
 \Psi_{i,i+1,n}= \Psi_{i+1, i,n} &=& - \sigma^2 \left[\left(1+ \mr e^{-\Gamma \tau}\right)^2 +\frac{(1-\mr e^{- \Gamma \tau})\tau}{m\Gamma} \left(\bm{j}_n(\bm{s}_i) + \bm{j}_n(\bm{s}_{i+1})\right)\right],\\
 \Psi_{i,i+2,n}= \Psi_{i+2, i,n} &=& \sigma^2 \mr e^{-\Gamma \tau},
 \label{eq:Pcerrend}
\end{eqnarray}
\end{widetext}
where
\begin{equation}
  \bm{j}_n(\bm{s}) = \sum\limits_{n'}\left.\frac{\partial \bm{F}_n (\bm x)}{\partial \bm{x}_{n'}} \right|_{\bm{x} = \bm{s}_i}
\end{equation} 
stands for the sum over the columns of the Jacoby matrix of the force $\bm{F} (\bm x) = \left( F_1(\bm x), \ldots,  F_N (\bm x) \right)$ evaluated at point $\bm{s}_i$.
Formula~(\ref{eq:Pcerr}) provides the likelihood of the observed noisy trajectory $\mathcal T_s$ to be used in the Bayesian inference protocol. The fact that the covariance matrix of the detection noise $\mathbf \Psi_n$ has five non-zero diagonals means that $\log [\det \left(\mathbf{C} + \mathbf{\Psi}_n\right)]$ cannot be directly calculated using Eq.~(\ref{eq:logdetC}). 
Instead, this quantity can be evaluated using a modified procedure for numerically solving the system of equations~(\ref{eq:Csystem}). 
First, the matrix $\left(\mathbf{C} + \mathbf{\Psi}_n\right)$ is decomposed to lower and upper triangular matrices $\mathbf L$ and $\mathbf L^{T}$ using Cholesky decomposition, i.e., $\left(\mathbf{C} + \mathbf{\Psi}_n\right) = \mathbf {L L}^{T}$, where the superscript $T$ denotes matrix transposition. 
In our case, $\mathbf L$ has the band structure with only three non-zero diagonals. 
Therefore, 
\begin{equation}
 \log [\det \left(\mathbf{C} + \mathbf{\Psi}_n\right)] = 2\sum \log [\mathrm{diag}\,(\mathbf {L})] 
\end{equation} 
due to the fact that the determinant of a triangular matrix is a product of its diagonal elements.
Similarly, the sum involving $\left(\mathbf{C} + \mathbf{\Psi}_n\right)^{-1}$ in~(\ref{eq:Pcerr}) can be evaluated in analogy with~(\ref{eq:Csystem}), replacing $\left(\mathbf{C} + \mathbf{\Psi}_n\right)$ with $\mathbf {L L}^{T}$.

\section{Analysis of BEEPSIS performance}
\label{sec:simul}

\subsection{Harmonic oscillator}
\label{sec:simharm}

The performance of BEEPSIS was systematically tested on the well-understood case of damped stochastic harmonic oscillator~\cite{NorrelykkePRE11}. To this end, we numerically simulated the Langevin equation of a particle confined in a harmonic potential $U(x) = - \kappa_0 x^2/2$ oriented along the $x$-axis, with the stiffness $\kappa_0$ linked to the eigenfrequency of oscillations $\Omega_0$ as $\Omega_0=\sqrt{\kappa_0/m}$. The particle was coupled to a thermal bath, which provided both an external driving force with white-noise spectrum (characterized by an effective temperature $T_0$) and energy dissipation (characterized by the viscous damping rate $\Gamma_0$). For the sake of simplicity, we analyzed the simulated trajectories in the absence of detection noise. 

The standard analysis of the dynamics of damped stochastic harmonic oscillator is based on fitting the power spectral density (PSD) of oscillator's position to the analytical model~\cite{NorrelykkePRE11} 
\begin{equation}
P_{xx}(\omega) = \left(\frac{2 k_B T_0 \Gamma_0}{m}\right)\frac{1}{\left(\Omega_0^2 - \omega^2 \right)^2 + \Gamma_0^2 \omega^2},
\label{eq:Lorenz}
\end{equation} 
which provides the system parameters ($\Omega_0, \Gamma_0, T_0$), assuming the knowledge of $m$. 
We used this method as a benchmark, against which the results of BEEPSIS inference were compared.

In general, the performance of both BEEPSIS and PSD fitting depends not only on the actual values of $(\Omega_0$, $\Gamma_0$, $T_0)$, but also on the length and sampling frequency of the analyzed trajectory. In the following, we focus on the dependence of the accuracy and precision of parameter inference on the characteristics of the trajectory, while the values of system parameters are randomly sampled from intervals of experimentally accessible values.
In order to obtain reliable quantification of the performance of both inference protocols, we repeat the process of random parameter sampling $10^4$ times and simulate a particle trajectory for each selected set of parameters, using the discretized Langevin equation~(\ref{eq:fLEv}) and (\ref{eq:fLEx})~\cite{ErmakJCompPhys80}. Subsequently, we analyze the simulated trajectories by BEEPSIS, using minimization of Eq.~(\ref{eq:bsum}), as well as by fitting the PSD of the trajectory with Eq.~(\ref{eq:Lorenz}).
Finally, the performance of both inference methods is characterized using the distributions of the ratios of estimated to input values of the system parameters (REI) evaluated across all randomly chosen sets of $(\Omega_0$, $\Gamma_0$, $T_0)$. 

In the simulations, we fixed the particle radius to $a=100$~nm and density to $\rho = 2000\,\mathrm{kg\, m^{-3}}$, which corresponds to the mass of $m=8.4\times10^{-18}\ \mathrm{kg}$. Subsequently, we generated stochastic trajectories for typical sets of experimental parameters ($\Omega_0$, $\Gamma_0$, $T_0$) that randomly sample the oscillation frequency $\Omega_0/(2\pi)$ in the range of $10-200$~kHz, ambient pressure $p$ in the range of $1-10^{5}$~Pa (following~\cite{LiNatPhys11}, corresponding values of $\Gamma_0$ were calculated to be in the range of $40-1.23\times10^6$ \mr{s^{-1}}), and ambient temperature $T_0$ in the range of $10-1000$~K. 
$\Omega_0$ and $T_0$ were sampled using uniform random distributions, while the sampling of $p$ was uniform on the logarithmic scale. 
The above ranges of simulation parameters cover the full spectrum of oscillator behavior from the heavily overdamped regime, through the underdamped stochastic regime, to almost deterministic oscillations. 

\begin{figure*}
 \includegraphics[width=\textwidth]{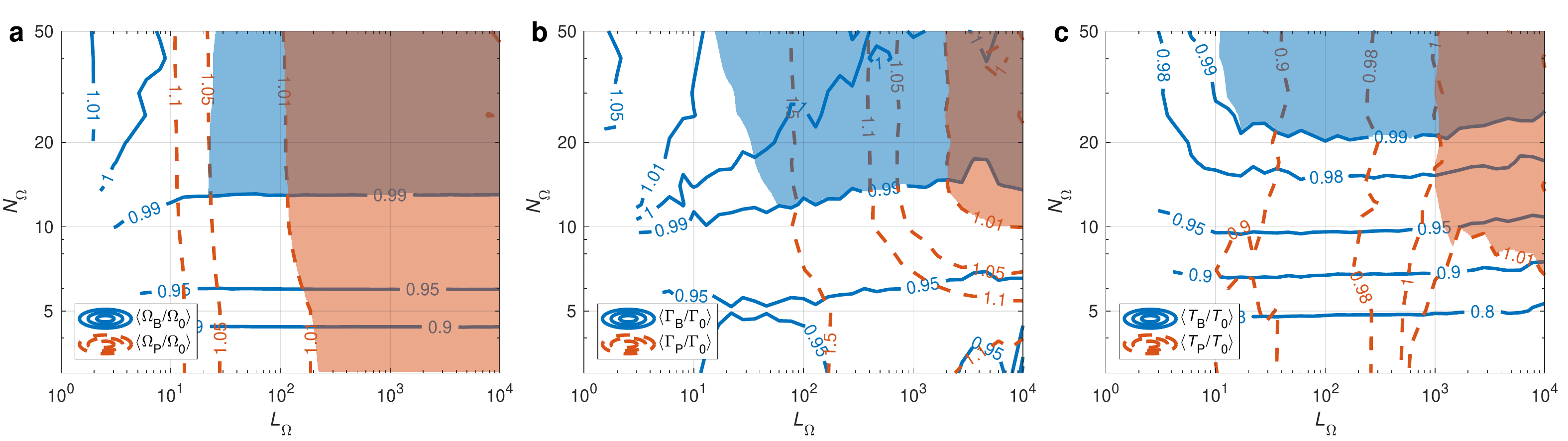}\\
 \includegraphics[width=\textwidth]{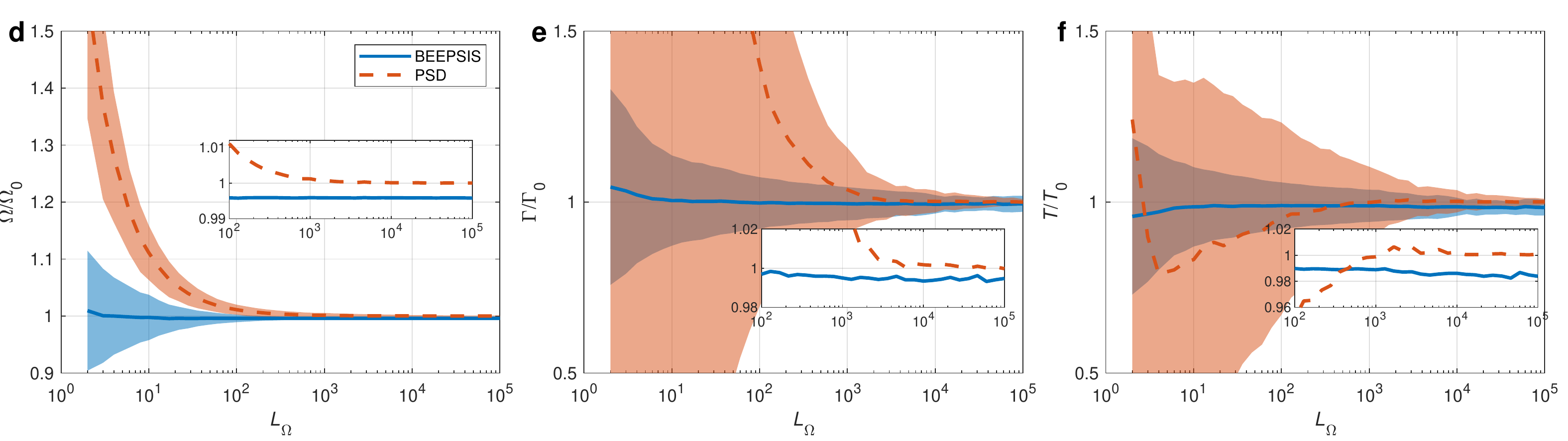}
 \caption{Comparison of performance of BEEPSIS and direct PSD fitting in estimating the parameters of a stochastic harmonic oscillator from simulated trajectories.  
 	(a--c): Dependence of the accuracy and precision of estimation of $(\Omega_0$, $\Gamma_0$, $T_0)$ on the number of acquired positions per oscillator period $N_{\Omega}$ (12 values logarithmically distributed in  the range of 3-50 points per period) and on the trajectory length $L_\Omega$ expressed as the number of oscillation periods (37 values logarithmically distributed in the range of  1-$10^4$ periods). For each combination of $N_{\Omega}$ and $L_\Omega$ (408 in total; 36 pairs of $(N_{\Omega},L_{\Omega})$ in the bottom left corner, where $N_{\Omega}$ and $L_{\Omega}$ are both small, were excluded from the analysis) we generated $10^4$ stochastic trajectories.
 	Each trajectory was simulated for a different random combination of input parameters $(\Omega_0$, $\Gamma_0$, $T_0)$, see text. 
 	For each trajectory, we estimated the values of system parameters ($\Omega_B,\Gamma_B,T_B$) using the BEEPSIS method, as well as the values of system parameters ($\Omega_P,\Gamma_P,T_P$) using the harmonic oscillator model~(\ref{eq:Lorenz}) that was fitted to the actual PSD of the trajectory.
 	Subsequently, we determined the accuracy and precision of the estimated parameters using the procedures described in the main text. 
 	The solid blue and dashed red contours depict the accuracy of parameter estimation by the BEEPSIS and PSD methods, respectively.
 	The blue and red shadings then mark the areas of both high accuracy and high precision (equivalent to low standard deviation -- std.) of inference: (a) $\langle\Omega_i/\Omega_0\rangle \in (0.99, 1.01)$, std. $\leq$ 0.02; (b) $\langle\Gamma_i/\Gamma_0\rangle \in(0.99,1.01)$, std. $\leq$  0.1; and (c) $\langle T_i/T_0\rangle \in (0.99, 1.01)$, std. $\leq$  0.1, with $i \in \{B, P \}$. 
	(d--f): Detailed dependence of the accuracy and precision of parameter estimation on the length of the trajectory $L_{\Omega}$ (1-$10^5$ periods) for fixed $N_{\Omega}=20$. 
	The blue and red curves show the accuracy of BEEPSIS and PSD inference, respectively, and the shadings of corresponding colors characterize the precision of both methods by delimiting the region of $\pm 1$ standard deviation. 
	Insets in (d-f) magnify the profiles of accuracy for $L_\Omega>100$. }
 \label{fig:HOsim}
\end{figure*}

Figure~\ref{fig:HOsim} summarizes the results of systematic analysis of the performance of both BEEPSIS and PSD fitting for various lengths and sampling frequencies of simulated trajectories. 
In order to correctly capture the dynamical evolution of the system trajectory for the given value of $\Omega_0$, the actual simulations were performed with a fine time step that corresponded to at least 100 points per period of particle's motion. Subsequently, we resampled the high-resolution trajectories with an effective sampling step $\tau$ giving the number of acquired positions per oscillation period $N_{\Omega} =2 \pi/(\tau \Omega_0) \in \langle 3,50\rangle$.
The length of the trajectory $L_{\Omega}$ was then expressed as a multiple of oscillation periods, $L_{\Omega}= L / N_{\Omega}\in\langle1, 10^4\rangle$. 
For each combination of $N_{\Omega}$ and $L_{\Omega}$, an independent set of input parameters ($\Omega_0$, $\Gamma_0$, $T_0$) was generated and an independent trajectory was simulated. 
Figures \ref{fig:HOsim} (a) - (c) illustrate the accuracy of parameter inference 
by BEEPSIS (solid blue contours; parameters identified by the subscript $B$) and PSD fitting (red dashed contours; parameters identified by the subscript $P$) as a function of $N_\Omega$ and $L_\Omega$. 
The accuracy $\langle \Omega_i/\Omega_0\rangle$ (Fig. \ref{fig:HOsim}a), $\langle \Gamma_i/\Gamma_0 \rangle$  (Fig. \ref{fig:HOsim}b), and $\langle T_i/T_0\rangle$ (Fig. \ref{fig:HOsim}c), with $i \in \{B, P \}$, is defined as the value of REI averaged over the full set of input system parameters ($\Omega_0$, $\Gamma_0$, $T_0$) used in the simulations for the particular trajectory length $L_\Omega$ and sampling frequency $N_\Omega$.
One can see that the accuracy of BEEPSIS is predominantly determined by the sampling rate $N_\Omega$. For $N_\Omega\gtrsim20$, BEEPSIS can reach $\pm1\%$ deviation from the actual input values of all three studied parameters for trajectories as short as $L_\Omega \sim 10$ periods. 
On the other hand, the accuracy of PSD fitting is mostly dominated by the trajectory length $L_\Omega$, whereas it is relatively insensitive to $N_\Omega$. In particular, accuracy of $\pm$1\% in estimating the oscillation frequency can be achieved for trajectory lengths of $L_\Omega\gtrsim 100$ periods with sampling rates of merely $N_\Omega =3$. However, the same level of accuracy in estimating the damping rate $\Gamma$ and effective temperature $T$ requires longer trajectories with finer sampling, $L_\Omega\gtrsim 10^3$ and $N_\Omega\gtrsim 10$.   

In order to obtain reliable estimates of the system's parameters, both accuracy and precision of the used inference procedure have to be sufficiently high.
The precision of inference can be quantified by the width of the distribution of REI observed for the particular length $L_\Omega$ and sampling frequency $N_\Omega$ of the simulated trajectory, which is proportional to the standard deviation (std.) of the corresponding REI ensemble. Specifically, high precision of inference is equivalent to small value of std.
In Figs.~\ref{fig:HOsim}(a) - (c), the regions of both high accuracy and high precision are marked by blue and red shading for BEEPSIS and PSD fitting inference, respectively. 
For all three parameters ($\Omega_0, \Gamma_0, T_0$), the colored shadings mark the regions of $\{L_\Omega, N_\Omega \}$ with the estimation accuracy of $\pm1\%$, i.e., $0.99<\langle$REI$\rangle<1.01$, whereas the high precision is defined by std. $<0.02$ for $\Omega_i/\Omega_0$ and by std. $<0.1$ for $\Gamma_i/\Gamma_0$ and $T_i/T_0$, with $i \in \{B, P \}$. 
The boundaries of the regions of high accuracy and high precision for the two inference techniques follow the previously discussed general trends: BEEPSIS performs well already for short trajectories, provided they are sufficiently well sampled, whereas PSD fitting needs longer trajectories with somewhat relaxed sampling demands.

As discussed above, BEEPSIS requires at least 20 points per period in order to achieve both good accuracy and precision of inference. 
This is primarily caused by the initial requirement that the force acting on the particle is constant over two consecutive time steps of the trajectory [see discussion below Eq.~(\ref{eq:fLE2})], which restricts the maximal permissible displacement of the particle in a single step. 
A deeper insight into the performance of BEEPSIS and PSD fitting inference can be gained from Figs. \ref{fig:HOsim}(d) - (f), which compare the accuracy (lines) and precision (shadings of corresponding colors) of both methods for a fixed $N_\Omega = 20$  and trajectory lengths $L_\Omega$ varying over the range extended up to $10^5$. 
With increasing trajectory length, BEEPSIS quickly reaches high levels of accuracy and precision in estimating the values of ($\Omega_0, \Gamma_0, T_0$) that are only attainable with PSD fitting inference based on trajectories about two orders of magnitude longer. 
For trajectories with $L_\Omega > 10^4$, the precision of parameter estimation is comparable for both inference methods.  
However, BEEPSIS slightly underestimates the values of all three parameters, with $\langle$REI$\rangle<1$ even for the longest considered trajectories.
This bias is approximately $\sim$ 0.5 \% in estimating the frequency $\Omega_0$ and viscous damping rate $\Gamma_0$ and about 3-times higher ($\sim$ 1.5 \%) in estimating the temperature $T_0$. 
As indicated in Figs.~\ref{fig:HOsim}(a) - (c), the bias can be reduced (or even completely eliminated) by increasing further the sampling frequency $N_\Omega$. Therefore, we conclude that it is again related to the degree of validity of the assumption of a constant force in two subsequent steps of the trajectory.

In Appendix~\ref{app:simho}, we present additional analysis of the dependence of BEEPSIS performance on the damping strength of the harmonic oscillator, quantified by the damping ratio $2\Omega_0/\Gamma_0$ that defines the transition from the underdamped ($2\Omega_0/\Gamma_0 > 1$) to the overdamped ($2\Omega_0/\Gamma_0 < 1$) regime. 

\subsection{Duffing oscillator}
\label{sec:Duffing_sim}
In addition to the case of harmonic oscillator, we analyzed the performance of BEEPSIS using simulated trajectories of a small particle confined in an anharmonic trap with Duffing force profile, $F_{\mathrm{D}}(x) = - m \Omega_0^2 x \left(1-\xi x^2\right)$, where $\xi > 0$ is the strength of Duffing nonlinearity. 
Such a force is typically found in optical tweezers formed by a tightly focused laser beam and results from the approximately Gaussian transverse profile of optical intensity within the beam focal region, which deviates from the idealized parabolic profile (see also Section~\ref{sec:experiments_OT}). The third-order term in $F_{\mathrm{D}}(x)$ leads to the softening of the trap for larger displacements from the equilibrium position, which may have a strong influence on the dynamics of motion of optically levitated particles \cite{GieselerNatPhys13,FlajsmanovaSR20}.
The systematic analysis of the accuracy and precision of inference of the Duffing oscillator parameters was carried out following the procedure described in the previous section. 
In this case, the simulated trajectories were only processed with BEEPSIS.
The results presented in Appendix \ref{app:simduff} illustrate that BEEPSIS represents a robust tool for inference of parameters of nonlinear systems that cannot be readily characterized by explicit models. 

\section{Experimental results}
\label{sec:experiments}
We will now demonstrate the applicability of BEEPSIS inference to the quantitative characterization of stochastic motion of nanoparticles confined in force fields generated by focused laser beams. 
Optically levitated nanoparticles have become an indispensable experimental tool to study the stochastic dynamics under a wide range of ambient conditions (see, e.g., \cite{Millen_RPP_2020,Gieseler_entropy_2018} and references therein). 
Due to the possibility of dynamic shaping of the light intensity and/or phase distribution by spatial light modulators, it is possible to create various types of reconfigurable force landscapes, which allow to experimentally study and verify many fundamental theoretical concepts, such as transition rates across a potential barrier \cite{KramersPHYS40,ReimannPhysRep02,Rondin2017}, generation of non-classical quantum states \cite{NeumeierArxiv22}, or various types of light-matter coupling \cite{RieserScience22}. 

In the following, we will apply BEEPSIS to three experimentally relevant cases of increasing complexity: 
\begin{enumerate}
 \item A nanoparticle confined in anharmonic optical tweezers,
 \item Particle motion in a double-well potential with a time-dependent force component for excitation of transitions across the potential barrier,
 \item Far-from-equilibrium system of a particle orbiting around the center of a circularly-polarized optical trap due to the non-conservative spin force. 
\end{enumerate}
Each of these cases requires a different treatment by BEEPSIS and may serve as a starting point for various other experimental configurations. 
The first case represents a simple one-dimensional (1D) problem, in which the force profile can be parameterized by an analytical formula. 
In the second case, which is still 1D, the force profile cannot be directly described by a single analytical function; instead, a smooth spline interpolation is used to characterize the spatial variation of the force. 
Moreover, a time-dependent force component (in the simple harmonic form) is present.
In both cases 1 and 2, all the parameters of the system - including the damping rate and temperature - can be inferred even when the input particle trajectories are influenced by detection noise. 
Finally, in the last considered case, the two-dimensional (2D) profile of the force field contains a non-conservative component, which induces sustained orbital motion around the center of the optical trap.
Furthermore, the force field is reconstructed from multiple measurements carried out at different trapping powers, with no prior knowledge of its spatial dependence.

\subsection{Optical tweezers}
\label{sec:experiments_OT}
\begin{figure*}
  \includegraphics[width=\textwidth]{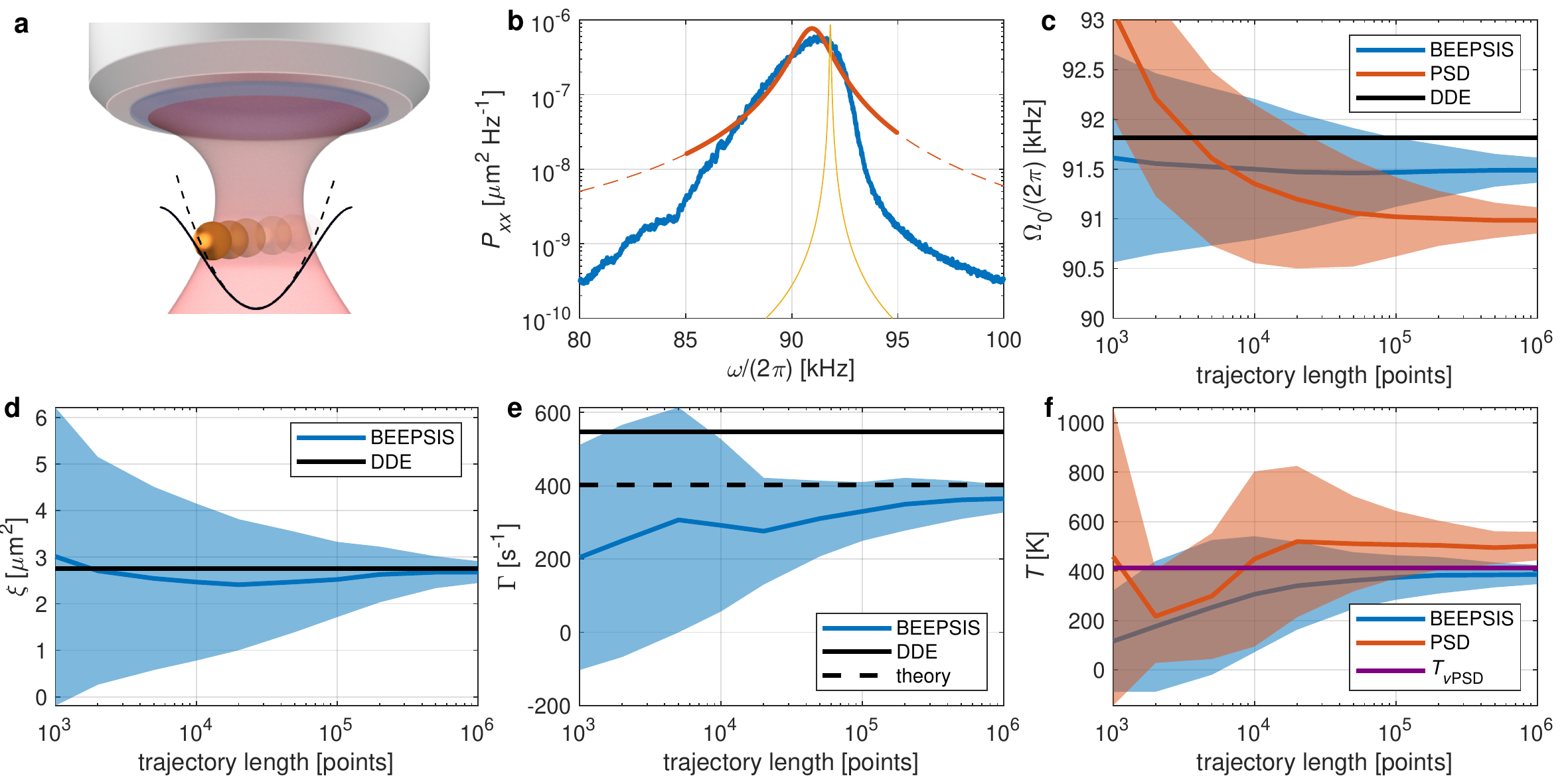}
  \caption{Quantitative characterization of particle motion in single-beam optical tweezers. (a) Illustration of a particle trapped in optical tweezers with anharmonic and harmonic potential profiles (solid/dashed curves). 
  (b) Power spectral density (PSD) obtained from the recorded trajectory of the optically trapped particle [blue curve], fit of the experimental PSD by the harmonic oscillator spectrum~(\ref{eq:Lorenz}) [red curve; the solid section depicts the spectral interval included in the fit, the dashed section is the extension of the harmonic model to other frequencies], and harmonic PSD calculated for the experimental oscillation eigenfrequency $\Omega_0/(2\pi) = 91.8$ kHz and viscous damping rate $\Gamma$ theoretically predicted for the given particle radius and ambient pressure. Temperature was arbitrarily adjusted to fit the vertical range of the plot [yellow curve].
  (c-f) Dependence of the accuracy and precision of estimation of the system parameters $\{\Omega_0,\xi,\Gamma,T\}$ by the BEEPSIS, PSD fitting, and DDE/velocity PSD methods on the trajectory length. For each estimated parameter, only the applicable inference methods are depicted. The thick curves show the mean values of the parameters for the given trajectory length, the shadings of corresponding colors characterize the inference precision by delimiting the region of $\pm 1$ standard deviation. (c) Oscillation eigenfrequency $\Omega_0$; (d) strength of Duffing nonlinearity $\xi$; (e) viscous damping rate $\Gamma$; (f) effective kinetic temperature of the particle motion $T$.
  Experimental parameters: particle radius $a = 85$ nm, ambient pressure 10 Pa, trapping power $P \simeq 100$ mW, position sampling rate 1.765 MHz (approximately 19.2 points per period).}
  \label{fig:exp1}
\end{figure*}

A nanoparticle confined and manipulated with a single focused laser beam that forms so-called {\em optical tweezers} [see Fig.~\ref{fig:exp1}(a) for a schematic illustration] represents a simple, yet extremely versatile experimental tool for controlled investigation of stochastic processes, with numerous applications ranging from cell and molecular biology through microrheology of complex fluids to surface and colloidal chemistry~\cite{GieselerAOP2021}. In addition, such a nanoparticle levitated in vacuum can be used as the starting point in experiments that attempt to cool its thermal center-of-mass motion toward the quantum ground state~\cite{MagriniNAT21,TebbenjohannsNAT21}. 

An optically trapped nanoparticle is usually described as a damped harmonic oscillator, with the optical force acting as a linear spring with stiffness $\kappa_0$ pulling the particle back into the trap center. 
However, as the particle moves farther from its equilibrium position in the trap, one may observe deviations from the ideal harmonic force profile. 
The most prominent anharmonic term is the Duffing-type cubic non-linearity, which - in the case of optical tweezers - softens the effective stiffness $\kappa$ of the optical trap with increasing particle displacement~\cite{GieselerNatPhys13,YonedaJPB17,FlajsmanovaSR20}. 
Specifically, the optical force along a single characteristic direction of optical tweezers can be described as 
\begin{equation}
 F_{\mathrm{D}}(x) = -m\Omega_0^2 x \left(1-\xi x^2\right),
 \label{eq:duff}
\end{equation} 
where $\Omega_0 = \sqrt{\kappa_0/m}$ is the harmonic angular eigenfrequency of the trap and $\xi>0$ is the strength of Duffing nonlinearity (see also discussion in Section~\ref{sec:Duffing_sim}). 
Due to the presence of a nonlinear term in the force profile~(\ref{eq:duff}), inference of parameters of stochastic Duffing oscillators is significantly more involved than the analysis of their harmonic counterparts~\cite{FlajsmanovaSR20}.

Our quantitative characterization of nonlinear optical tweezers was based on experimentally recorded trajectories of a silica nanoparticle (nominal radius $a = 85$ nm, density $\rho = 2000\ \mathrm{kg\,m^{-3}}$; Bangs Laboratories) trapped in vacuum (pressure 10 Pa) by a laser beam ($\lambda = 1064$ nm, trapping power at the particle location $P \simeq 100$ mW; Mephisto, Coherent) focused by an aspheric lens with NA = 0.7. Full details of the experimental setup are provided in~\cite{FlajsmanovaSR20}.
The particle trajectories were recorded using a quadrant photodiode (QPD) with the sampling frequency of 1.765 MHz, while the oscillation frequency of the particle, $f_0 = \Omega_0/(2 \pi)$, was $\sim$ 91.8 kHz [see the blue curve in Fig. \ref{fig:exp1}(b), which shows the experimental PSD]; this gave us approximately 19.2 points per one period of oscillation. According to the results of Section \ref{sec:simharm} and Appendix \ref{app:simduff}, with this sampling rate, the parameters inferred by BEEPSIS should not be influenced by any significant bias. 
In total, 60 consecutive trajectories of length $10^6$ points each were measured. 
Figure \ref{fig:exp1}(b) shows the PSD of the particle position obtained using all recorded trajectories combined together (the blue curve). 
We can clearly see that the strongly asymmetric shape of the experimental PSD peak is not well represented by the fitted spectral profile of a harmonic oscillator~(\ref{eq:Lorenz}), indicated in Fig. \ref{fig:exp1}(b) by the red curve. 
For comparison, the yellow curve shown in Fig. \ref{fig:exp1}(b) is the PSD profile of a pure harmonic oscillator with the resonant frequency set to the experimental value of $\Omega_0/(2\pi) = 91.8$ kHz, damping rate predicted for the used particle radius and ambient pressure using the theory derived in~\cite{LiNatPhys11}, and temperature arbitrarily adjusted to fit the peak height into the vertical range of the plot. 
The yellow curve directly demonstrates the change of the PSD peak shape and the associated spectral broadening due to the presence of the non-linear Duffing force. 

In order to evaluate the performance of BEEPSIS for various lengths of experimental trajectories, in analogy to Figs. \ref{fig:HOsim}(d) - (f), we divided the actual measured trajectories into shorter segments of lengths ranging from 10$^3$ points (60$\times10^3$ segments in total) up to $10^6$ points (60 segments in total). 
For each of these trajectory segments, we then applied BEEPSIS to estimate the parameters of the nonlinear optical trap $\Omega_0$, $\xi$ [see Eq.~(\ref{eq:duff})], as well as the viscous damping rate $\Gamma$ and temperature $T$ associated with the ambient atmosphere.
Using the procedure outlined in section \ref{sec:error}, we included the effect of detection uncertainty in the inference. To this end, we determined the standard deviation of the position detection error $\sigma$ from the high-frequency noise floor of the experimental PSD as $\sigma = 2.57$ nm (all experimental trajectories combined together) and inserted this value of $\sigma$ into the transition probability of noisy trajectories~(\ref{eq:Pcerr}) used in the likelihood minimization. 
As a benchmark of BEEPSIS performance in estimating the characteristics of the force profile $\Omega_0$, $\xi$ and damping rate $\Gamma$, we adopted the analysis of ensemble-averaged transient trajectories by the numerical solution of deterministic Duffing equation -- the DDE method, see~\cite{FlajsmanovaSR20} -- which processed the whole ensemble of 60 experimental trajectories at once. 
For formal comparison, we also analyzed the PSDs of individual trajectory segments with varying length, which were independently fitted with Eq.~(\ref{eq:Lorenz}) to obtain the values of $\Omega_0$ and $T$. 
As discussed above, harmonic trap profile associated with~(\ref{eq:Lorenz}) does not represent well the actual analyzed experimental data. Nevertheless, we find it instructive to include this method of inference in the comparison to quantitatively illustrate the systematic biases associated with using the idealized linear model to describe a realistic nonlinear optical trap.

Figures \ref{fig:exp1}(c) - (f) compare the performance of BEEPSIS and the other two inference methods mentioned above. 
For all analyzed parameters, the precision of BEEPSIS inference increased with increasing trajectory length; this is directly reflected in the gradual narrowing of blue-shaded regions in which the variation of parameter ensembles estimated from BEEPSIS for the given trajectory length lies within $\pm1\, \mathrm{std.}$ 
For trajectories shorter than $\sim5\times10^3$ points, only $\Omega_0$ could be determined with a good precision, whereas the standard deviations of the estimates of $\xi$, $\Gamma$, and $T$ were bigger than the mean values of these parameters. 
These experimental results coincide with the results of simulations presented in Section \ref{sec:simharm} and in Appendices \ref{app:simho} and \ref{app:simduff}.
Therefore, in the quantitative comparison of the various inference methods, we focus on the results obtained for the longest studied trajectory length of $L=10^6$ points ($\approx5\times10^4$ periods).  

The estimated oscillation eigenfrequency $\Omega_0$ as a function of the trajectory length is plotted in Fig. \ref{fig:exp1}(c), in which the blue, red, and black curves show the mean values of $\Omega_{0, \mathrm{BEEPSIS}}$, $\Omega_{0, \mathrm{PSD}}$, and $\Omega_{0, \mathrm{DDE}}$ obtained from BEEPSIS, PSD fitting, and DDE, respectively. 
Since the DDE method only uses the full trajectory length of $10^6$ points for the estimation, its prediction appears as a constant in the plot. 
The estimates of $\Omega_0$ obtained from trajectories with $10^6$ points using all three inference methods are then summarized on the first line of Table \ref{tbl:twres}. 
As indicated by Fig. \ref{fig:exp1}(c), the value of $\Omega_{0, \mathrm{PSD}}$ reaches its maximum for the shortest trajectory segments. This can be intuitively explained by the fact that for the majority of these segments, the particle moved close to the trap center where the softening of the trap due to the Duffing nonlinearity is negligible~\cite{FlajsmanovaSR20}.
In addition, Table \ref{tbl:twres} shows that the values of $\Omega_{0, \mathrm{BEEPSIS}}$ and $\Omega_{0, \mathrm{DDE}}$ are slightly higher than $\Omega_{0, \mathrm{PSD}}$.
This can be expected, as the PSD model given by Eq.~(\ref{eq:Lorenz}) does not include the Duffing softening and, consequently, $\Omega_{0, \mathrm{PSD}}$ represents an effective value for all possible particle locations within the anharmonic potential profile of the optical tweezers.
Regarding the reference value obtained by the DDE method, it follows that $\Omega_{0,\mathrm{BEEPSIS}} \simeq 99.6 \% \Omega_{0, \mathrm{DDE}}$, which is consistent with the slight underestimation of model parameters by BEEPSIS that was predicted by the stochastic simulations, see Appendix \ref{app:simduff}.

Figure \ref{fig:exp1}(d) shows the dependence of the strength of Duffing nonlinearity $\xi$ obtained by BEEPSIS and DDE (blue curve and black line, respectively) on the trajectory length. Because the harmonic oscillator model that leads to the PSD~(\ref{eq:Lorenz}) does not feature $\xi$, PSD fitting inference was not applicable for this parameter.
The values of $\xi$ extracted from trajectories with $10^6$ points, which are reported on the second line of Table \ref{tbl:twres}, show that the BEEPSIS and DDE estimates agree within the uncertainty of inference.

\begin{table}
 \caption{Characteristic parameters of motion of a nanoparticle confined in optical tweezers, determined from experimental trajectories of the particle using BEEPSIS, PSD fitting, and DDE/velocity PSD methods of inference. Theoretical value of $\Gamma$ was calculated from the model presented in~\cite{LiNatPhys11}. Experimental parameters: particle radius $a = 85$ nm, ambient pressure 10 Pa, trapping power $P \simeq 100$ mW, trajectory length $10^6$ points.}
 \label{tbl:twres}
 \begin{ruledtabular}
 \footnotesize
 \begin{tabular}{rr@{ $\pm$}lr@{ $\pm$}lr@{}lc}   
   & \multicolumn{2}{c}{BEEPSIS} & \multicolumn{2}{c}{PSD} & \multicolumn{2}{c}{DDE/$T_{v \mathrm{PSD}}$} & theory\\
   \hline
   $\Omega_0/(2\pi)$ [kHz] & 91.50& 0.13 & 91.0 & 0.1 & 91.819&$\pm$ 0.009&--\\
   $\xi\ \mathrm{[\mu m ^{-2}]}$ & 2.69&0.09 &\multicolumn{2}{c}{--}  & 2.76&$\pm$ 0.03&--\\
   $\Gamma\, \mathrm{[s^{-1}]}$& 370&14& \multicolumn{2}{c}{(11$\pm$2)$\times10^3$}& 548&$\pm$2& 406.6\\
   $T$ [K] &390&14&500 & 60 & 414 & &--\\
 \end{tabular}
\end{ruledtabular}
\end{table}

The viscous damping rate $\Gamma$ obtained by BEEPSIS and DDE inference (blue curve and black line, respectively) is depicted in Fig. \ref{fig:exp1}(e) and its respective values $\Gamma_{\mathrm{BEEPSIS}}$, $\Gamma_{\mathrm{DDE}}$ corresponding to the longest trajectories with $10^6$ points are given on the third line of Table \ref{tbl:twres}.
These values are close to the theoretical prediction of $\Gamma_{\mathrm{th}} = 406.6\ \mathrm{s^{-1}}$ determined for the experimental particle radius and ambient pressure from the model reported in~\cite{LiNatPhys11} [see also the black dashed line in Fig. \ref{fig:exp1}(e)]; $\Gamma_{\mathrm{BEEPSIS}}$ actually agrees with $\Gamma_{\mathrm{th}}$ within the range of inference uncertainty.
On the other hand, the values of $\Gamma_{\mathrm{PSD}}$ obtained by PSD fitting [not shown in Fig. \ref{fig:exp1}(e)] are $\sim 20 \times$ higher than all three of $\Gamma_{\mathrm{BEEPSIS}}$, $\Gamma_{\mathrm{DDE}}$, and $\Gamma_{\mathrm{th}}$. 
This is caused by the fact that the peak in the PSD of experimental trajectories is broadened due to the nonlinear Duffing effect; thus, the fitted value of $\Gamma_{\mathrm{PSD}}$ reflects these nonlinear shifts of the trapping eigenfrequency rather than the damping of the particle's motion. 
As a result, fitting of experimentally obtained PSD to the harmonic oscillator model of Eq.~(\ref{eq:Lorenz}) is unsuitable for characterizing the damping rate of the Duffing oscillator.

Finally, the effective kinetic temperatures $T_{\mathrm{BEEPSIS}}$ and $T_{\mathrm{PSD}}$ of the particle motion inferred by BEEPSIS and PSD fitting are plotted in Fig. \ref{fig:exp1}(f) [blue and red curve, respectively]. 
The values of $T_{\mathrm{BEEPSIS}}$ and $T_{\mathrm{PSD}}$ estimated from trajectories with $10^6$ points are then summarized on the last line of Table \ref{tbl:twres}. We compare them with the reference value $T_{v{\mathrm{PSD}}}$ calculated by applying the equipartition theorem to the integrated area below the velocity PSD determined from the full set of experimental trajectories~\cite{HebestreitRSI18} [see also the purple horizontal line in Fig. \ref{fig:exp1}(f)]. 
This reference is chosen instead of the DDE method \cite{FlajsmanovaSR20} as it does not require any parametrization and gives a direct temperature value without any fitting. 
As argued above, the PSD profile~(\ref{eq:Lorenz}) does not follow the actual spectral distribution of energy [compare the blue and red curve in Fig. \ref{fig:exp1}(b)]. Consequently, PSD fitting largely overestimates the effective ambient temperature. 
Simulations presented in Appendix \ref{app:simduff} indicate that $T_{\mathrm{BEEPSIS}}$ is likely underestimated by $\sim$2\%. At the same time, $T_{v{\mathrm{PSD}}}$ might be overestimated due its strong sensitivity to the high-frequency detection noise~\cite{HebestreitRSI18}. 
Therefore, we may expect the real experimental value to lie between $T_{\mathrm{BEEPSIS}}$ and $T_{v{\mathrm{PSD}}}$.

\subsection{Double-well potential and time-dependent force}
\label{sec:experiments_DW}
Tailoring the intensity or phase profiles of the trapping light beams by spatial light modulators has been extensively used in the overdamped regime of optical trapping for the past 25 years~\cite{Grier2003,GieselerAOP2021}. 
In the case of optically confined stochastic systems with inertia, beam shaping has been applied, for example, to dynamically control optically bound matter~\cite{SvakOptica21}, or to design a procedure for creating non-Gaussian quantum states of massive dielectric objects~\cite{NeumeierArxiv22}.
One of the simplest and experimentally most widely adopted profiles of optical forces is based on the double-well potential with two potential minima close to each other, separated by a barrier of a well-defined, controllable height. 
Depending on the relationship between the mean energy of the trapped particle and the barrier height, the particle can remain confined in the vicinity of a single potential minimum, transition back and forth between the two wells with a characteristic rate determined by the actual configuration of the system, or oscillate freely across the full extent of the double-well potential, virtually unimpeded by the barrier.
Optical double-well potential can serve as a model system to systematically study phenomena ranging from thermally activated escape from a metastable state~\cite{KramersPHYS40,Rondin2017}, through stochastic resonance \cite{GammaitoniRMP98}, to amplification of mechanical motion~\cite{NeumeierArxiv22}. 
In general, quantitative description of all these phenomena requires the precise knowledge of the spatial profile of the confining potential/force.

Most theoretical studies assume that the double-well potential can be described by a low-order polynomial (typically 4$^{\mathrm {th}}$-order). However, this ideal functional dependence is usually not achievable by the experimental techniques of beam shaping, in which the generated potential profiles have to conform to the principal limitations imposed by diffraction of light as well as by experimental uncertainties. 
Therefore, in order to provide a more realistic formal description for the analysis of experimental trajectories, we represent the force by a smooth spline profile that is defined by $M$ force values estimated at equidistantly separated grid points (knots), which span the full extent of the particle motion. 
One could, in principle, use a higher-order polynomial to describe the same force profile; however, with this approach, the minimization algorithm often requires data pre-processing in order to avoid problems with machine precision and the interpretation of the results might not be as straightforward. 

\begin{figure}
  \includegraphics[width=\columnwidth]{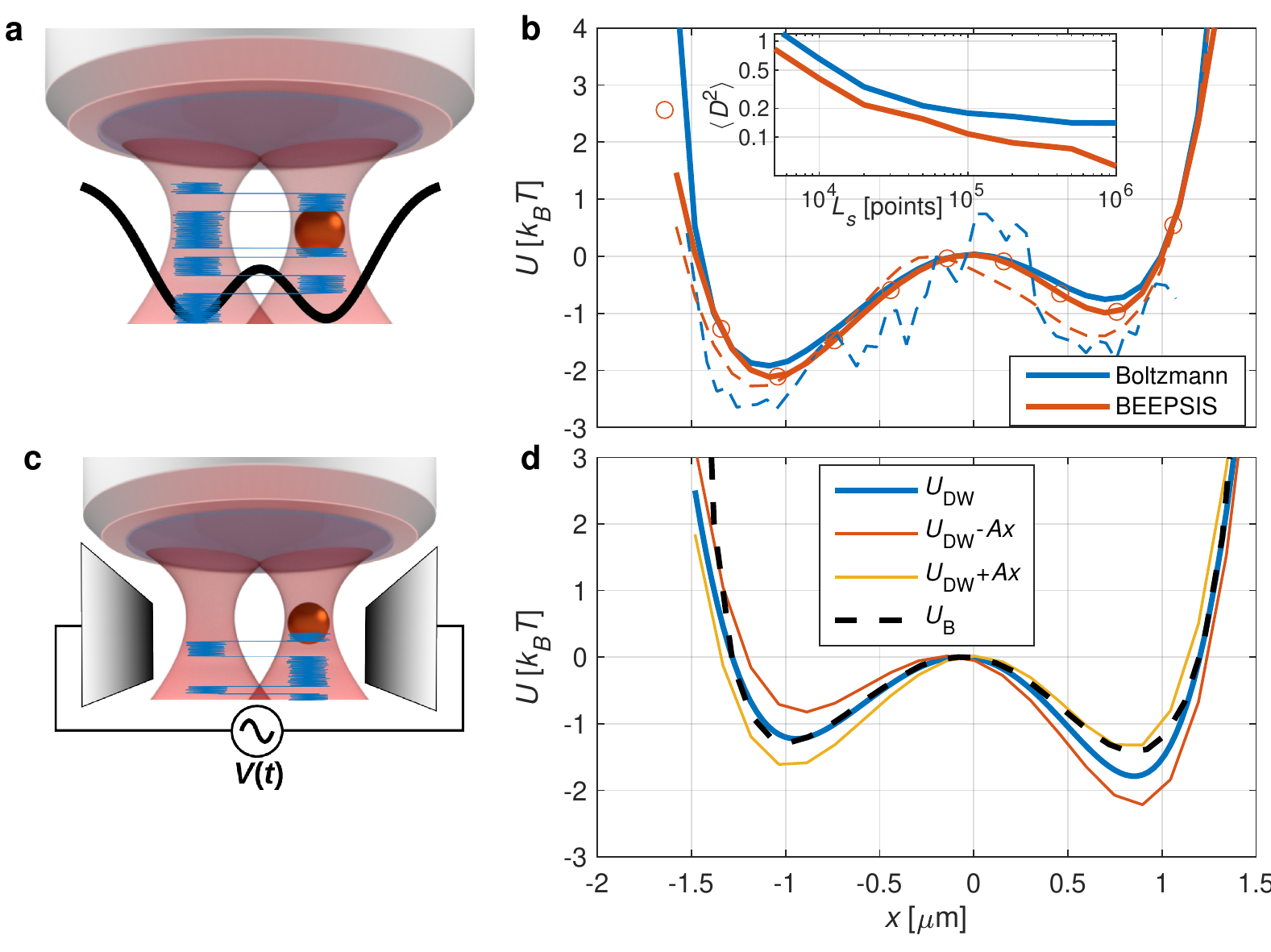}
  \caption{Quantitative characterization of particle motion in static and dynamic double-well potentials. (a) Illustration of a particle trapped in a static double-well optical potential with an example trajectory displaying random transitions between the two wells. (b) Comparison of double-well potential profiles reconstructed using stationary Boltzmann distribution (solid blue curve) and BEEPSIS (solid red curve) from an experimental trajectory with $10^7$ points. Red circles indicate locations where the force values were estimated by BEEPSIS. Dashed curves show the potential profiles reconstructed from the initial trajectory section of 5000 points. Inset shows the averaged squared distance $\langle D^2 \rangle$ between the potential profiles estimated from shorter and full trajectories, see Eq.~(\ref{eq:D_2}). (c) Illustration of a particle trapped in a double-well potential harmonically modulated by external electric field. 
  (d) Stationary component of the net confinement potential estimated by BEEPSIS (solid blue curve) and the extent of its changes during time-dependent harmonic modulation of the double-well trap depth (solid red and yellow curves). Dashed black curve represents the time-averaged net potential obtained from the Boltzmann distribution. For details, see Eq.~(\ref{eq:F_DW}) and related discussion in the text. 
  Experimental parameters: particle radius $a = 300$ nm, ambient pressure 1300 Pa, position sampling rate 200 kHz.
  } 
  \label{fig:dw}
\end{figure}

Figure \ref{fig:dw}(a) illustrates the motion of a particle confined in the double-well potential created by two partially overlapping optical traps, with an example of particle's trajectory randomly jumping back and forth between the two wells. 
Such motion can be treated as quasi 1D, oriented along the axis connecting the centers of the two wells.
Experimentally, the double-well potential was realized in a holographic optical tweezers system with counter-propagating laser beams~\cite{SvakOptica21}.
A silica nanoparticle (nominal radius $a = 300$ nm, density $\rho = 2000 \,\mathrm{kg\, m^{-3}}$; Bangs Laboratories) bearing electric charge was trapped by two pairs of adjacent counter-propagating laser beams 
(wavelength 1550 nm, total trapping power 85 mW; seed laser NKT Koheras ADJUSTIK X15 PM, fiber amplifier NKT Koheras BOOSTIK, beam shaping by digital micromirror device ViaLux V650L SuperSpeed V-Module)
at pressure of (1300 $\pm$ 400) Pa. 
Particle's motion was simultaneously monitored by a fast CMOS camera (Phantom V611, Vision Research) and by a QPD, with the sampling rate set to 200 kHz for both devices. Following the data acquisition, the non-linear QPD signal was calibrated using the particle's trajectory obtained from the camera record. The total length of the final processed trajectory recorded by the QPD was $10^7$ points.

Initially, we analyzed the full recorded trajectory using the Boltzmann probability distribution (BD)~\cite{GieselerAOP2021}, which allowed us to express the potential profile in the units of $k_B T$ [see the solid blue curve in Fig.~\ref{fig:dw}(b)]. 
Upon the assumption of the particle moving at the ambient temperature $T = 293$ K and with the particle's mass calculated from its nominal radius and density, we then numerically differentiated the potential to obtain the force at $M=11$ points distributed equidistantly over the full extent of the particle's motion. 
These force values were subsequently used as the initial guess for the BEEPSIS inference.
The potential profile recovered by BEEPSIS is depicted as the solid red curve in Fig. \ref{fig:dw}(b), with the red circles marking the points (spline knots) where the values of the optical force were estimated. 
We can see that the overall Boltzmann and BEEPSIS potential profiles correspond fairly well to each other (see below for a comment on the remaining observed differences). 
The BEEPSIS-inferred values of the effective temperature and viscous damping rate were $T_{\mathrm{eff}} = ( 213.22 \pm 0.02)\ \mathrm{K}$ and $\Gamma = (17642 \pm 2) \ \mathrm{s^{-1}}$, respectively. 
As the negative likelihood function minimized by BEEPSIS depends only on the ratio $T_{\mathrm{eff}}/m$, the fact that the estimate of $T_{\mathrm{eff}}$ is lower than the actual ambient temperature $T = 293$ K is most likely caused by the particle's real mass being larger than the expected value based on the nominal diameter and density of the particle.  
Specifically, the observed ratio of $T$ and $T_{\mathrm{eff}}$ implies that the actual particle radius is $\sim \sqrt[3]{T / T_{\mathrm{eff}}} \approx 1.12 \times$ the nominal value, which is within the range specified by the manufacturer. 
Taking into account the adjusted value of the particle radius, the theory predicts the viscous damping rate of $\Gamma_{\mathrm{theory}} = 14078 \ \mathrm{s^{-1}}$ at $T = 293$ K and the experimentally measured ambient pressure of 1300 Pa~\cite{LiNatPhys11}. 
This calculated $\Gamma_{\mathrm{theory}}$ is lower than the inferred value; however, we can obtain a nearly perfect match upon adjusting the pressure to 1600 Pa, which lies within the measurement uncertainty of the experimentally used pressure gauge.
Possible uncertainty in the determination of $T_{\mathrm{eff}}$ might have caused the small residual difference between the BEEPSIS and Boltzmann potential profiles, namely, the greater depth of the potential wells of the former profile.

Similar to the parameter inference from the experimental data obtained with optical tweezers, we reduced the length of the processed trajectory to just 5000 initial points and analyzed this short trajectory by both BD [dashed blue curve in Fig. \ref{fig:dw}(b)] and BEEPSIS [dashed red curve]. 
We can see that for the BD estimation, the reconstructed potential profile exhibits numerous random oscillations and is generally shifted towards more negative values with deeper wells, in comparison to the profile obtained from the full trajectory with $10^7$ points. 
In contrast, even with a short processed trajectory, BEEPSIS correctly reconstructed the profile of the deeper (left-hand side) potential well. 
The profile of the right-hand side potential well was not reconstructed correctly by either method, as the particle moved through this region less often within the analyzed trajectory segment.
The inset of Fig.~\ref{fig:dw}(b) quantifies the quality of the potential profile reconstruction by both methods as a function of the partial trajectory length $L_s$.
For this analysis, the whole trajectory of $L=10^7$ points was divided into segments of length $L_s$.
Each segment was then independently analyzed using the procedures described above and the average squared distance between the potential curves was calculated as 
\begin{equation}
  D^2 = \frac{1}{\Delta x} \int \left( U_s - U_{\mathrm{full}}\right)^2 \mathrm d x.
  \label{eq:D_2}
\end{equation} 
Here, $\Delta x$ is the full extent of the particle's motion along the $x$-axis (which also defines the region of integration) and $U_s$, $ U_{\mathrm{full}}$ are the potential profiles recovered from the partial trajectories and from the full trajectory, respectively, using either BD or BEEPSIS inference. 
The curves plotted in the inset of Fig.~\ref{fig:dw}(b) show the values of $\langle D^2 \rangle$ averaged over all trajectories of the given length $L_s$.
Comparison of the plots then reveals that for all $L_s$, BEEPSIS gives $1.2 - 3\times$ better correspondence between the potential curves determined from the short and long trajectories. 

In addition to studying a static 1D double-well potential, we extended the measurement protocol by applying time-dependent, spatially uniform electric field parallel to the potential's $x$-coordinate to the same trapped particle. 
This electric field tilted the net potential profile experienced by the electrically charged trapped particle and, consequently, induced particle transitions between the neighboring potential wells. 
To apply the field, we connected harmonically varying voltage $V(t)$ from a function generator to a pair of quasi-planar electrodes placed in the experimental vacuum chamber in the vicinity of the trapping region [see Fig.~\ref{fig:dw}(c) for illustration].
The nominal frequency of the applied harmonic voltage was 250 Hz; however, we observed a slow phase drift of this voltage over long time periods. 
In order to avoid uncertainties in inferring the instantaneous electric force applied to the particle, we independently measured $V(t)$ and used this signal to express the net space- and time-dependent force $F_{\mathrm{DW}}(x,t)$ acting on the particle as
\begin{equation}
F_{\mathrm{DW}}(x,t) = F_{\mathrm{DW}}^{\mathrm{stat}}(x) + A V_n(t),
\label{eq:F_DW}
\end{equation}
where $F_{\mathrm{DW}}^{\mathrm{stat}}(x)$ is the double-well optical confining force and $A V_n(t)$ represents the time-dependent force component, which is proportional to the measured voltage $V_n(t)$ normalized to the interval $\langle -1, 1 \rangle$.
The static as well as the dynamic force components were analyzed using the previously described approach. 
Figure \ref{fig:dw}(d) shows the static potential profile obtained solely from the stationary force component $F_{\mathrm{DW}}^{\mathrm{stat}}(x)$ (solid blue curve), as well as the instantaneous potential profiles at times when the maximal positive/negative electric field was applied, which are associated with the net forces $F_{\mathrm{DW}}^{\mathrm{stat}}(x) \pm A$ (solid red and yellow curves). 
All three potential profiles were obtained by integrating the corresponding forces along the $x$-axis.
For comparison, we also show the BD-based potential [dashed black curve in Fig. \ref{fig:dw}(d)] obtained from the whole trajectory, i.e., ignoring the time-dependent electric force acting on the particle. 
While this time-averaged Boltzmann potential profile is close to the stationary part of the time-varying potential inferred by BEEPSIS, some differences ($\sim$ 0.4 $K_B T$) are visible close to the right-hand side well. 
This difference might be caused by the time-dependent force that brings the particle out of thermal equilibrium and enhances its transition rates from the slightly deeper right-hand side well. 
Details of this process will be the subject of a future study.
The inferred values of the viscous damping rate and effective temperature are $\Gamma = (17870 \pm 2) \ \mathrm{s^{-1}}$ and $T_{\mathrm{eff}} = ( 229.72 \pm 0.03)\ \mathrm{K}$, respectively, which is comparable to the values obtained in the case of a stationary double well. 
Therefore, the exposure of the particle to a time-varying force field did not introduce any biases into the estimation of the environmental characteristics. 

The above findings demonstrate the versatility of BEEPSIS in the quantitative characterization of complex potential landscapes with time-varying force profiles. Specifically, we have shown that BEEPSIS provides access not only to the spatial profile of the force, but also to the environmental parameters (effective temperature and viscous damping rate), which cannot be readily determined by other conventional methods (e.g., Boltzmann distribution analysis) in this type of underdamped dynamical system. 
Moreover, we have provided evidence that - in contrast to the Boltzmann distribution approach - BEEPSIS can reliably estimate the force profiles even from relatively short trajectories.

\subsection{Spin force -- Externally driven stochastic system far from equilibrium}
\label{sec:experiments_spin}
\begin{figure*}
	\includegraphics[width=\textwidth]{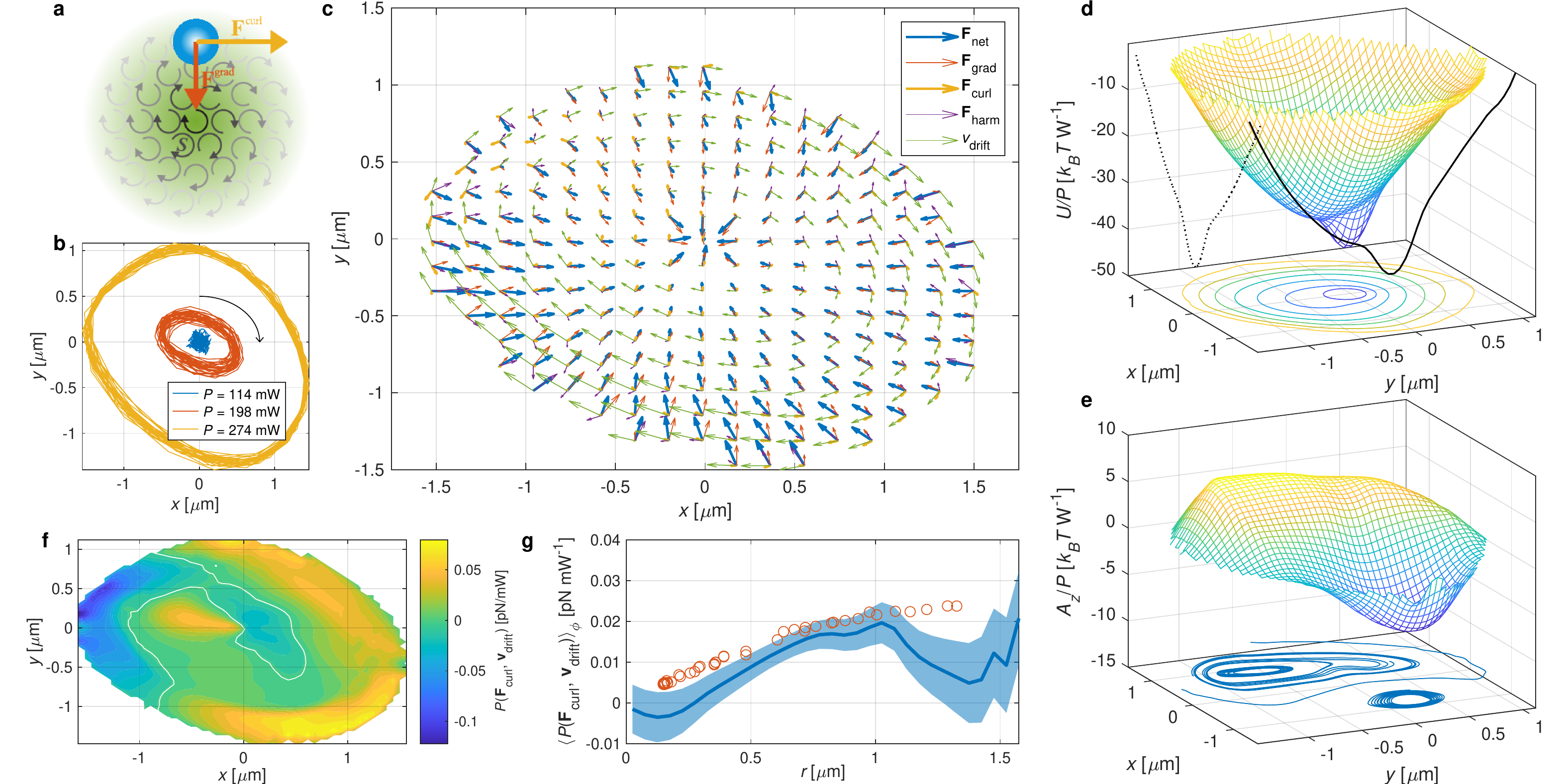}
	\caption{Quantitative characterization of particle motion in conservative and non-conservative 2D force fields generated within a circularly polarized optical trap. (a) Illustration of the origin of the azimuthal spin force in an inhomogeneous, circularly polarized light beam. $\bm{S} = \Im\{\bm{E}^{*} \times \bm{E}\}$: local optical spin field in the transverse $xy$-plane of the beam, $\bm{F}_{\mathrm{grad}}$: radial conservative gradient force, and $\bm{F}_{\mathrm{curl}}$: azimuthal non-conservative spin force acting on a particle positioned off the beam axis. 
	(b) Short segments of three particle trajectories recorded for different trapping powers: $P = 198, 274$ mW (red and yellow curves) -- orbiting cases, $P = 114$ mW (blue curve) -- non-orbiting case. 
	(c) Power-normalized 2D profiles of the net optical force $\bm{F}_{\mathrm{net}}$ recovered using BEEPSIS (blue arrows) and of its conservative (red arrows), non-conservative (yellow arrows), and harmonic (purple arrows) components $\bm{F}_{\mathrm{grad}}$, $\bm{F}_{\mathrm{curl}}$, and $\bm{F}_{\mathrm{harm}}$ obtained using Helmholtz-Hodge decomposition. 
	Further, the green arrows show the direction and magnitude of the power-normalized average drift velocity $\bm{v}_{\mathrm{drift}}$.
	(d) Power-normalized scalar potential $U$ associated with $\bm{F}_{\mathrm{grad}}$ (meshed surface), its 2D projection on the $xy$-plane (contours), and 1D profiles at the intersections with planes $x=0$ (black solid curve) and $y=0$ (black dotted curve). 
	(e) Power-normalized $z$-component of the vector potential $\bm{A}$ (meshed surface) and the streamlines of $\bm{F}_{\mathrm{curl}}$ in the $xy$-plane (contours).  
	(f) Signed magnitude of the projection $P\left(\bm{F}_{\mathrm{curl}},\bm{v}_{\mathrm{drift}}\right)$ of  $\bm{F}_{\mathrm{curl}}$ on the local direction of $\bm{v}_{\mathrm{drift}}$, normalized to the trapping power. White contour separates the regions with positive (accelerating) and negative (decelerating) values of $P\left(\bm{F}_{\mathrm{curl}},\bm{v}_{\mathrm{drift}}\right)$. 
	(g) Azimuthal average $\langle P\left(\bm{F}_{\mathrm{curl}},\bm{v}_{\mathrm{drift}}\right) \rangle_{\phi}$ as a function of the particle distance from the trap center $r$ (solid blue curve). Blue shading indicates the standard deviation of $\langle P\left(\bm{F}_{\mathrm{curl}},\bm{v}_{\mathrm{drift}}\right) \rangle_{\phi}$ and the points marked by red circles are the values of the azimuthal force reported in \cite{SvakNC18}. 	
	Experimental parameters: particle radius $a = 770$ nm, ambient pressure 400 Pa, position sampling rate 250 kHz.}
	\label{fig:orbit}
\end{figure*}

For a long time, extraordinary spin component of the Poynting momentum of light has been considered to be merely a virtual quantity, formally required by Belifante's symmetrization of the canonical stress energy tensor to conserve angular momentum~\cite{Soper2008Classical}. 
Recently, however, it has been experimentally shown that this transverse spin momentum has a physical manifestation in a non-conservative force acting on an object placed into an inhomogeneous circularly polarized light beam~\cite{Antognozzi2016Direct,SvakNC18}. 
In particular, we have experimentally demonstrated that this non-conservative force component may induce orbiting of a particle confined in a circularly polarized optical trap along a closed quasi-circular trajectory~\cite{SvakNC18}. 
The origin of this force is schematically depicted in Fig. \ref{fig:orbit}(a). Intuitively, it arises due to the non-negligible transverse gradient of the local optical spin $\bm{S}$ across the particle diameter, which generates an unbalanced net pushing force $\bm{F}_{\mathrm{curl}}$ in the azimuthal direction.
As reported in~\cite{SvakNC18}, the orbiting behavior only appears above a certain threshold value of the trapping power where the centripetal gradient force $\bm{F}_{\mathrm{grad}}$ balances the inertial centrifugal force caused by the orbiting motion and, at the same time, the non-conservative azimuthal force $\bm{F}_{\mathrm{curl}}$ is balanced by the viscous drag acting on the particle.

In the experimental demonstration and quantitative characterization of the transverse spin force, we used a silica particle (nominal radius $a = 770$ nm, density $\rho = 2000 \,\mathrm{kg\, m^{-3}}$; Bangs Laboratories) whose 2D trajectories were recorded with the sampling rate of 250 kHz by a QPD (InGaAs G6849; Hamamatsu). Prior to the measurement, the QPD position signal was calibrated against the trace obtained simultaneously with a fast CMOS camera (Phantom V611, Vision Research).
The particle was confined in vacuum in a counter-propagating Gaussian beam trap whose polarization state could be controlled using quarter-wave plates on either side of the trap (see~\cite{SvakNC18} for the full description of the experiment). 
Mapping of the azimuthal spin force was then carried out with circularly polarized beams of opposite handedness, whereas linearly polarized beams were used in the measurement of the viscous damping rate.
In total, 205 and 142 trajectories of length 1~s were recorded in circularly and linearly polarized traps, respectively. In these experiments, laser power $P$ was varied in the range between 60 -- 320 mW, covering values of $P$ both below and above the threshold for orbiting, and the ambient pressure was fixed at 400 Pa.
Examples of particle trajectories observed in the circularly polarized trap for three different values of $P$ are provided in Fig. \ref{fig:orbit}(b).
The blue trajectory localized near the coordinate origin corresponds to the system state before the onset of the orbiting motion, while the red and yellow trajectories demonstrate orbiting states. 
One can see that for each laser power, the particle trajectory covers only a small fraction of the area over which we would like to reconstruct the full 2D force profile. 
In order to obtain the desired force map over the $xy$-plane, we analyzed the trajectories in the following manner:
\begin{enumerate}
  \item  Using particle trajectories recorded in the linearly polarized trap, we first determined the value of the viscous damping rate $\Gamma = (11700\pm700)\ \mathrm{s^{-1}}$  by fitting the harmonic trap model, see Eq. ~(\ref{eq:Lorenz}), to the PSD of the measured trajectory. This value of $\Gamma$ served as a fixed parameter in the subsequent BEEPSIS inference.
  \item For each trajectory recorded with the same particle in the circularly polarized trap, we applied BEEPSIS in 2D to extract the force profile $\bm{F}_{\mathrm{net}}(x,y)$ on a rectangular grid of $(49\times50)$ points that covered the extent of particle motion in both $x$ and $y$ directions (\emph{local} grid).
  We assumed a general 2D force profile, with the force vector being approximately constant over spatial bins centered around the grid points.
  Grid spacing was selected to reconstruct the force profiles with a reasonably fine spatial resolution while still keeping good precision even in the rarely visited bins for all trajectories recorded both below and above the orbiting threshold.
  The force field was directly obtained using the fast procedure described in Appendix \ref{app:binning}, which does not require the full numerical minimization of the negative log-likelihood.
  \item  In addition to the forces, we also determined the mean drift velocities $\bm{v}_{\mathrm{drift}}(x,y)$ of the particle at each bin of the local grid (calculated using the central difference) and recorded the total number of times the trajectory passed through the given bin.
  \item For each analyzed trajectory, we normalized the vector fields of the force and drift velocity by the actual used trapping laser power $P$.
  \item Finally, we combined together the results obtained for all individual trajectories.  
  First, we created a regular {\em global} grid of points covering the total extent of the particle motion observed in the whole experimental series with varied trapping power 
  and embedded all local grids into the global one.
  Subsequently, the power-normalized values of the force and drift velocity determined on the local grids were assigned to the bins of the global grid where they were averaged. 
  In calculating the averages, we weighted the contribution of each local grid by the number of passages of its corresponding trajectory through the given bin of the local grid. 
\end{enumerate}

The resulting power-normalized vector fields of the net force $\bm{F}_{\mathrm{net}}$ and drift velocity $\bm{v}_{\mathrm{drift}}$,
obtained by merging the results of analysis of 205 trajectories, are plotted in Fig. \ref{fig:orbit}(c) as the sets of blue and green arrows, respectively. 
The force field points mostly to the trap center; however one may also observe a weak force component oriented in the perpendicular direction. 
Similarly, the arrows depicting the drift velocity point in the direction of the orbiting motion. 
However, the directions of drift do not exactly follow circular trajectories, as they would in the ideal theoretical case with full rotational symmetry~\cite{SvakNC18}; instead, the particle orbits along deformed trajectories, which was also directly observed in the experiments. 

The observed asymmetries in the profiles of $\bm{F}_{\mathrm{net}}$ and $\bm{v}_{\mathrm{drift}}$ indicate that a simple decomposition of the net force field to the radial and azimuthal components would not lead to proper determination of the spin-related force. 
Therefore, in order to decompose the force field into the conservative and non-conservative components, we applied the Helmholtz-Hodge decomposition (HHD) \cite{BhatiaIEEE13} which separates any general force field $\bm{F}_{\mathrm{net}}(x,y)$ into three parts
\begin{equation}
	\begin{split}
 	\bm{F}_{\mathrm{net}}(x,y) = & \,\,-\grad U(x,y) + \curl \bm{A}(x,y)  \\
 	&\,\,+ \bm{F}_{\mathrm{harm}}(x,y).
	\end{split}
	\label{eq:HHD}
\end{equation} 
Here, the first component on the right-hand side is the conservative force $\bm{F}_{\mathrm{grad}}$ described by the scalar potential $U$ (also called curl-free or irrotational), the second component is the non-conservative divergence-free force $\bm{F}_{\mathrm{curl}}$ described by the vector potential $\bm{A}$ (also called incompressible or solenoidal), and the third component is the so-called harmonic force with zero divergence and curl.
The first two terms come from the classical Helmholtz theorem of the vector calculus.
The additional harmonic force is then identically zero for decomposition carried out on infinite space, but is non-zero on bounded decomposition regions, where the boundary conditions of the HHD require that the irrotational field is perpendicular to the boundary, while the incompressible field is tangential \cite{BhatiaIEEE13}.   
The HHD can be performed by numerically solving the Poisson partial differential equation.
The results of this procedure are plotted in Fig. \ref{fig:orbit}(c), which shows all three components of the force vector field ($\bm{F}_{\mathrm{grad}}$, $\bm{F}_{\mathrm{curl}}$, and $\bm{F}_{\mathrm{harm}}$) as red, yellow, and purple arrows, respectively. 
At most locations, the conservative force (red) points towards the center of the trap and acts as a centripetal force. 
The non-conservative component (yellow) then represents the spin force accelerating the particle motion along its closed orbits. Thus, it is mostly colinear with the drift velocity (green arrows). 
The most notable differences are in the locations of the top-left quadrant (negative $x$ and positive $y$) where the spin force actually acts against the direction of the drift velocity and slows down the sustained motion of the particle.
This leads to the deformations of the particle trajectory that can be seen in Fig. \ref{fig:orbit}(b). 

The reconstructed scalar potential $U$ and vector potential $\bm{A}$ are depicted in Figs. \ref{fig:orbit}(d,e), respectively. 
Please note that since the force field is only 2D, lying in the $xy$-plane, the vector potential has only one component, $A_z$, in the $z$-direction. 
The scalar potential $U$, shown in Fig.~\ref{fig:orbit}(d), displays an asymmetric shape along the $y$ direction with a tight, localized well that gradually spreads to a wider spatial extent. Changing curvature of the potential profile then indicates varying stiffness of particle confinement. 
The localized central potential well mentioned above resembles the profile of a conventional optical trap with linear polarization in which the confined particle does not experience any extraordinary forces induced by the light polarization state. 
However, as the particle moved out of the central region during its orbital motion at higher trapping powers, the shape of the potential well changed from the ideal parabolic (or even Gaussian) one. 
The reconstructed shape of the potential profile likely reflects the actual distribution of the light intensity influenced by the spatial light modulator and by diffraction effects in the optical path, with the asymmetry in the $y$-direction caused by tiny misalignment of the counter-propagating light beams.

The vector potential $A_z$, depicted in Fig. \ref{fig:orbit}(e), exhibits a symmetrically located pair of a maximum and a minimum. 
In an analogy with the fluid flow, this means that the incompressible component of the force field $\bm{F}_{\mathrm{curl}}$ features two vortices with opposite senses of rotation and the streamlines of the vector field (depicted as a projection in the $xy$-plane) flow in two opposite directions in the $xy$-plane. 
This agrees with the above stated observation that -- especially in the top-left coordinate quadrant -- the non-conservative force field acts against the orbiting motion of the particle. 
This phenomenon can be even more clearly seen in the calculated projection $P\left(\bm{F}_{\mathrm{curl}},\bm{v}_{\mathrm{drift}}\right)$ of the non-conservative force upon the direction of the drift velocity, see Fig. \ref{fig:orbit}(f).
Here, the $xy$-plane is separated into two regions with positive and negative values of $P\left(\bm{F}_{\mathrm{curl}},\bm{v}_{\mathrm{drift}}\right)$ that represent acceleration or deceleration of the orbiting motion. 
The region with the negative projection on the non-conservative force is predominantly located in the top-left corner of the map shown in Fig. \ref{fig:orbit}(f). However it also extends toward the trap center, reflecting the asymmetries related to the alignment of the trapping beams.

Finally, we calculated the azimuthal average $\langle P\left(\bm{F}_{\mathrm{curl}},\bm{v}_{\mathrm{drift}}\right) \rangle_{\phi}$ as a function of the particle distance from the trap center $r$ [see Fig.~\ref{fig:orbit}(g)]. 
To this end, we evaluated $P\left(\bm{F}_{\mathrm{curl}},\bm{v}_{\mathrm{drift}}\right)$ on concentric rings with radii $r$ in the $xy$-plane and calculated the average value (solid blue curve) and standard deviation (blue shading) of $P\left(\bm{F}_{\mathrm{curl}},\bm{v}_{\mathrm{drift}}\right)$ along the rings. 
This quantity can be compared to the values of the azimuthal force reported in~\cite{SvakNC18}, which are depicted in Fig.~\ref{fig:orbit}(g) as red circles. 
Each data point taken from~\cite{SvakNC18} represents the azimuthal force component obtained from a single measurement at a fixed trapping power under the assumption of a perfectly circular trajectory of radius $r$. 
As illustrated by Fig.~\ref{fig:orbit}(g), the values obtained in~\cite{SvakNC18} and in this work by BEEPSIS display qualitative and quantitative correspondence, even though the used methods of data processing were completely different. 
The differences between BEEPSIS and Ref. \cite{SvakNC18} are most pronounced for small radii of the orbiting motion. 
Here, BEEPSIS even predicts that the particle should orbit in the reverse direction. 
However, in reality, the particle moves along a distorted non-circular trajectory; hence, $\langle P\left(\bm{F}_{\mathrm{curl}},\bm{v}_{\mathrm{drift}}\right) \rangle_{\phi}$ calculated along a fixed-radius circle does not fully represent the complex experimental motion of the particle. In addition, rather coarse grid spacing close to the trap center might distort the BEEPSIS results in this region due to the averaging of force profiles originating from multiple trajectories recorded at different trapping powers, mixing together the diffusive and orbiting modes of particle's motion. 
On the other hand, the correspondence of the two methods of analysis is very good for particle distances from the trap center between 0.5--1 $\mathrm{\mu m}$. 

The systematic study of the azimuthal spin force acting on an optically trapped particle, presented in this section, illustrates the applicability of BEEPSIS to quantitative inference of parameters of non-stationary systems that are out of thermal equilibrium. 
By merging the results of multiple independent measurements carried out at different trapping powers, which were subsequently properly normalized, we were able to reconstruct the extended 2D vector force field containing both conservative and non-conservative components, without making any simplifying assumptions about the nature and/or symmetry of the field.
Furthermore, we mapped the transition from the diffusive motion of the confined particle, exhibited close to the center of the optical trap, to the stable orbiting motion observed for larger displacements of the particle from the origin. 
We compared selected results of BEEPSIS inference (in particular, the dependence of the magnitude of the non-conservative force on the distance from the trap center) with the data obtained using alternative methods of analysis~\cite{SvakNC18} and found a good qualitative and quantitative agreement.

\section{Conclusions}
\label{sec:conclusion}
We proposed a novel inference method for quantitative characterization of force fields, viscous damping rate, and effective ambient temperature acting on microscopic stochastic systems whose dynamics is influenced by inertia:  Bayesian Estimation of Experimental Parameters in Stochastic Inertial Systems (BEEPSIS). This method can be applied to a general dynamical system, possibly out of thermal equilibrium, subject to an external force field containing conservative, non-conservative, or time-varying components. Starting from the full formal solution of the discretized second-order Langevin equation, the likelihood of the actual observed trajectory of the system was first expressed solely as a function of directly measurable quantities (i.e., system’s positions in discrete times) and the sought parameters of the local force field and the ambient environment (damping rate, temperature). These parameters were then found as the most likely values compatible with the observed trajectory. In its most generalized formulation, BEEPSIS is capable of explicitly including the effects of the position measurement error into the likelihood formula. 

First, we systematically tested the validity of BEEPSIS inference, using simulated trajectories of model stochastic harmonic oscillators and anharmonic Duffing oscillators whose parameters were varied over experimentally relevant values of oscillation eigenfrequency, viscous damping rate, ambient temperature, and (for the Duffing oscillator) strength of nonlinearity. Moreover, we compared the performance of BEEPSIS against the results of fitting the power spectral density (PSD) of oscillator positions to the theoretical model, which represents an established method of quantitative characterization of harmonic oscillators~\cite{NorrelykkeRSI2010}. This systematic analysis provided the accuracy and precision of BEEPSIS inference as functions of the length and sampling frequency of simulated trajectories. On the basis of its results, we identified the main requirements on the trajectories that are necessary to achieve sufficient correspondence between the simulation inputs and estimated parameters. In particular, we found that the position sampling rate plays the most critical role, requiring at least 20 measured points per single characteristic period of the studied trajectory to obtain unbiased, precise estimates. When the condition of the minimal sampling rate is met, BEEPSIS can reach high levels of accuracy and precision in estimating the values of the oscillation eigenfrequency, viscous damping rate, and ambient temperature even from trajectories as short as ten periods. In contrast, a comparable performance is only attainable with PSD fitting inference based on trajectories about two orders of magnitude longer. 

Subsequently, we applied BEEPSIS to the quantitative characterization of stochastic motion of nanoparticles levitated in vacuum in optical fields with tailored intensity profiles. Such optically levitated nanomechanical systems represent a unique, largely tunable experimental testbed that is ideally suited for systematic studies of stochastic dynamics under controlled environmental conditions. The analysis of experimentally recorded trajectories was carried out for three cases with increasing complexity, namely, a nanoparticle confined in anharmonic optical tweezers, a nanoparticle moving in a double-well potential with a time-dependent force component, and an out-of-equilibrium system of a nanoparticle stably orbiting around the center of a circularly-polarized optical trap due to the non-conservative spin force. Each of these cases highlighted a different aspect of quantitative inference in stochastic inertial systems, serving as a model example for other related experimental configurations. Specifically, in the optical tweezers case, the 1D force profile could still be described by a closed analytical formula, whereas in the double-well case, a smooth spline interpolation was used to characterize the complex 1D spatial profile of the force, which, moreover, varied in time. The last considered case then involved a 2D force field containing a non-conservative component with a priori unknown spatial dependence. Whenever possible, we compared the results of BEEPSIS inference with the corresponding values of system parameters provided by alternative methods of trajectory analysis that served as performance benchmarks (PSD fitting~\cite{NorrelykkeRSI2010}, DDE method~\cite{FlajsmanovaSR20}, Boltzmann distribution~\cite{FlorinAPA98}, and orbiting equation for the azimuthal spin force~\cite{SvakNC18}). This comparison revealed that BEEPSIS could render the system parameters with a good accuracy and precision already from segments of trajectories that were significantly shorter than those required by the alternative tested strategies for inference. Furthermore, BEEPSIS provided simultaneous access to parameters that would otherwise have to be determined by analyzing multiple independent measurements (for example, effective temperature and viscous damping rate in the double-well potential characterized by the static Boltzmann distribution approach).

BEEPSIS is based on the second-order Langevin dynamics observed in the presence of a general force field that, in principle, can be non-conservative or time-varying and whose spatial dependence does not have to be analytically described or a priori known. Therefore, BEEPSIS inference can also be applied to out-of-equilibrium and non-stationary systems that are not amenable to the standard treatment based on the premise of thermal equilibrium. The main technical limitation of the current implementation of BEEPSIS lies in the assumption of vanishingly small changes of the acting force between two consecutive points of the measured trajectory, which imposes restrictions on the minimal sampling rate needed for unbiased estimation of the system parameters. Replacing the requirement of constant force with a weaker assumption of linear step-to-step variations of the force profile would relax the limits on the necessary sampling rate, which, in turn, would directly enable quantitative experimental characterization of faster dynamical events. One of the most promising directions of further development of the BEEPSIS inference framework is its extension to stochastic inertial systems subject to active external cooling of their thermal motion, which are among the hottest candidates for experimental testing of quantum-mechanical laws on the mesoscopic scale. The cooling schemes that are most widely used in the field of optical levitation – parametric cooling~ \cite{Gieseler2012Subkelvin} and cold damping~ \cite{TebbenjohannsPRL19}) – are based on electronic feedback loops that monitor the state of motion of the levitated particle and subsequently apply an effective dissipative force opposing the motion. Random noise added to the detected position of the particle enters the feedback loop where it gets amplified; consequently, it acts as an effective stochastic source driving the particle’s motion independently from the ambient thermal bath. The correlation between the detector noise and particle’s dynamics, which has to be properly accounted for in the inference scheme, presents a major challenge. However, our results suggest the feasibility of including the effects of noise in the BEEPSIS treatment, expanding further its utility as a powerful tool for quantitative data analysis and a guide for experimentalists in choosing appropriate measurement conditions to achieve desired accuracy and precision.

\begin{acknowledgments}
  The research was supported by projects of the Czech Science Foundation (GF21-19245K), the Czech Academy of Science -- Praemium Academiae (AP2002) and The Ministry of Education, Youth and Sports.
    (CZ.02.1.01/0.0/0.0/16\_026/0008460).
\end{acknowledgments}

\appendix
\section{Covariance of misfits}
\label{app:ito}
The covariance matrix $\mathbf{C}_{ij} = \langle \bm{m}_i \bm{m}_j \rangle$ introduced in Eq.~(\ref{eq:Pc}) describes correlations of misfits at different times. 
For the purpose of evaluating $\mathbf{C}$, we can replace the expressions for misfits defined by Eq.~(\ref{eq:misfit}) by the stochastic integrals that appear on the right-hand side of Eq.~(\ref{eq:fLE2}), i.e. 
\begin{widetext}
\begin{equation}
 \bm{m}_i = \frac{1}{m\Gamma} \left\{ \int_0^\tau \left[1-e^{-\Gamma (\tau-t')} \right]\bm \xi (t_i+t') \mr d t' + e^{-\Gamma \tau} \int_0^\tau \left[e^{\Gamma t'} -1\right]\bm \xi (t_{i-1}+t') \mr d t'\right\}.
 \label{eqa:misfit}
\end{equation}
We will now demonstrate the calculation of the diagonal elements $\mathbf{C}_{ii}$ of the covariance matrix. Using~(\ref{eqa:misfit}), we can write
\begin{eqnarray}
\nonumber
 \mathbf{C}_{ii}& =& \frac{1}{m^2\Gamma^2} \left\langle  \left\{ \int_0^\tau \left[1-e^{-\Gamma (\tau-t')} \right]\bm \xi (t_i+t') \mr d t' + e^{-\Gamma \tau} \int_0^\tau \left[e^{\Gamma t'} -1\right]\bm \xi (t_{i-1}+t') \mr d t'\right\}^2 \right\rangle\\
 \nonumber
 &=& \frac{1}{m^2\Gamma^2} \left\langle  \left\{ \int_0^\tau \left[1-e^{-\Gamma (\tau-t')} \right]\bm \xi (t_i+t') \mr d t'\right\}^2 \right\rangle\\
 \nonumber
 && + \frac{1}{m^2\Gamma^2} \mr e^{-2\Gamma \tau}\left\langle \left\{  \int_0^\tau \left[e^{\Gamma t'} -1\right]\bm \xi (t_{i-1}+t') \mr d t' \right\}^2 \right\rangle\\
 && + \frac{2}{m^2\Gamma^2} \left\langle  \left\{ \int_0^\tau \left[1-e^{-\Gamma (\tau-t')} \right]\bm \xi (t_i+t') \mr d t' \right\}\times\left\{ e^{-\Gamma \tau} \int_0^\tau \left[e^{\Gamma t'} -1\right]\bm \xi (t_{i-1}+t') \mr d t'\right\} \right\rangle
 \label{eqa:Cii}
\end{eqnarray} 
\end{widetext}
The stochastic integrals in~(\ref{eqa:Cii}) can be  written in Stieltjes form as 
\begin{equation*}
 \int_0^\tau A(t') \xi (t_i+t') \mr d t' = \int_0^\tau A(t') \mr d B_{t'}
\end{equation*} 
where $B_{t'}$ represents 1D Brownian motion \cite{Risken,OksendalBook}. 
The variance (second moment) of the stochastic integrals can then be evaluated using It\^o isometry \cite{OksendalBook},
\begin{equation*}
 \left\langle \left\{ \int_0^\tau A(t') \mr d B_{t'} \right\}^2\right\rangle\! = 2k_BT m\Gamma \left\langle  \int_0^\tau A^2(t') \mr d t' \right\rangle\!,
\end{equation*} 
where we assumed the time covariance of Brownian motion given by Eq.~(\ref{eq:LEcov}), in contrast to the unit covariance used in~\cite{OksendalBook}.
With this result, the sum of the first two integrals in~(\ref{eqa:Cii}) leads to the value of coefficient $a$ [see Eq. (\ref{eq:Ca})].
The last term of~(\ref{eqa:Cii}) is identically zero due to the fact that the random variable integrals are evaluated over different time intervals starting at $t_{i}$ and $t_{i-1}$. 

The same procedure can also be adopted to evaluate the non-diagonal covariance elements $\mathbf{C}_{ij}$ with $i \neq j$. In this case, the only combinations of $i$ and $j$ that lead to non-zero ensemble averages of relevant stochastic integrals involved in the calculation fulfill $j = i \pm 1$. This observation then directly leads to the off-diagonal coefficient $b$ given by Eq.~(\ref{eq:Cb}).

\section{Taylor expansion of misfits and covariances}
\label{app:taylor}
For short time steps (i.e., $\Gamma \tau \ll 1$), one may replace the exponential function $e^{-\Gamma \tau}$ in Eqs. (\ref{eq:misfit}), (\ref{eq:Ca}), and (\ref{eq:Cb}) by its second-order Taylor expansion.
Furthermore, taking into account the second-order discrete-time approximations for the instantaneous velocity and acceleration, $\bm{v_i} = (\bm{x}_{i+1}-\bm{x}_{i-1})/(2\tau)$ and $\bm{a_i} = (\bm{x}_{i+1} - 2\bm{x}_{i} +\bm{x}_{i-1})/\tau^2$, respectively, one may expand Eqs. (\ref{eq:misfit}), (\ref{eq:Ca}), and (\ref{eq:Cb}) up to the second order as 
\begin{eqnarray}
 \bm{m_i} &\approx& \left[\bm{a_i}  +\bm{v_i} \Gamma - \frac{\bm{F(x_i)}}{m}\right]\tau^2,\\
 a&\approx&  \frac43 \frac{k_BT \,\Gamma}{m} \tau^3 \left[ 1 -  \Gamma \tau + \frac35 \,\Gamma^2 \tau^2 \right],  \\
 b&\approx&  \frac13 \frac{k_BT \,\Gamma}{m } \tau^3 \left[ 1 - \Gamma \tau +  \frac{11}{20} \,\Gamma^2 \tau^2 \right].
\end{eqnarray}
From the above expansions, it follows that $a\simeq4b$ in the first order of $\Gamma\tau$, with $\sim5$\% difference appearing in the second order.

\section{Velocity-explicit conditional probability of particle transitions}
\label{app:velocpdf}
The transition probability $P(\mathcal T | \bm{\phi})$ of the trajectory $\mathcal T$, described by Eqs.~(\ref{eq:Pc}--\ref{eq:Cb}) in the main text, explicitly depends on particle positions only. This in turn introduces correlations between adjacent time steps that lead to a non-diagonal covariance matrix of misfits and increase the complexity of the inference process. 
However, when the full phase-space trajectory of the particle is known, i.e., both position and momentum (velocity) are independently accessible, it is also possible to express the transition probability in a simplified form, which can appreciably speed up the minimization procedure, especially for more complex models with many parameters. The approximate inferred values of the parameters may subsequently be used as initial guesses in minimization based on the full model given by Eqs.~(\ref{eq:Pc}--\ref{eq:Cb}), with the total data processing time decreased.

Full formal description of a second-order stochastic system requires the knowledge of the probability that the particle, initially located at the phase-space point $(\bm{x}_i, \bm{v}_i)$, moves to the final phase-space point $(\bm{x}_{i+1}, \bm{v}_{i+1})$ during a time interval $\tau$. 
One can, in principle, derive two different marginal probability density functions (PDFs) for the final state, dependent solely either on the final velocity  $\bm{v}_{i+1}$ or on the final position $\bm{x}_{i+1}$.
For the velocity-dependent case, the transition probability can be obtained by solving the Fokker-Planck equation~\cite{Risken} 
\begin{equation}
 \begin{split}
 & P_v\left(  \bm{v}_{i+1},\tau | \bm{x}_i, \bm{v}_i\right) =  \sqrt{\frac{m}{2\pi k_BT} \frac{1}{(1-\mr e^{-2\Gamma \tau})}} \\ 
  & \times \exp \left\{ - 
     \frac{\left[ \bm{v}_{i+1}\! -\! \bm{v}_i \mr e^{-\Gamma \tau}  - \frac{\bm{F}(\bm{x}_i)}{m\Gamma}\left( 1\!-\! \mr e^{-\Gamma \tau} \right)\right]^2}{  \frac{2k_BT}{m} \left(1-\mr e^{-2\Gamma \tau}\right)}\right\},
 \end{split}
\label{eq:Pv}
\end{equation} 
where we assumed that both the force and viscous damping rate are constant during the transition.
This probability explicitly depends only on the velocities $\bm{v}_i, \bm{v}_{i+1}$ at the initial and final time; implicit dependence on the initial position $\bm{x}_i$ is manifested solely through the acting force $\bm{F}(\bm{x}_i)$. 
As a consequence, when $\bm {F} = 0$, equation (\ref{eq:Pv}) reduces to the transitional probability of a freely diffusing particle.
As discussed in \cite{FerrettiPRX20}, an inference procedure based on this PDF, with velocities obtained by difference schemes, leads to estimation of a biased damping factor $\Gamma$ that corresponds to only $\sfrac23$ of the true value.

The marginal PDF for the final position $\bm{x}_{i+1}$ can be introduced following the procedure of Chandrasekhar as~\cite{ChandrasekharRMP43}
\begin{eqnarray}
 \nonumber &&
 P_x\left(\bm x_{i\!+\!1},\tau | \bm{x}_i, \bm{v}_i\right) = \sqrt{\frac{m \Gamma^2}{2\pi k_BT} \frac{1}{(2\Gamma \tau\! -\!3 +\! 4 \mr e^{-\Gamma \tau}\!\!-\!\mr e^{-2\Gamma \tau})}} 
 \\ \nonumber &&
 \exp \left\{ \vphantom{\int} \right. -\begin{aligned}[t]
                  &\left[ \bm x_{\!i+\!1} \!-\! \bm{x}_i \! -\! \frac{\bm{v}_i}{\Gamma}\left(1\!-\!\mr e^{-\Gamma \tau}\right) \right. \\ 
                  &\quad\left.  - \frac{\bm F( \bm x_i)}{m\Gamma^2}\left( \Gamma \tau \!-\!1 \!+\!\mr e^{-\Gamma \tau}  \right)\right]^2\\
               \end{aligned}   
   \\  && \qquad\left.  
   \times\left[\frac{2k_BT}{m \Gamma^2}\left(2\Gamma \tau -3 + 4 \mr e^{-\Gamma \tau} -\mr e^{-2\Gamma \tau}\right)\right]^{-1}
   \right\}.
\label{eq:Px}
\end{eqnarray}

In both cases, the PDFs at time $(t+\tau)$ depend only on the phase-space coordinates at time $t$ and describe a Markov stochastic process.
Therefore, either of them could be adopted to express the probability of the whole trajectory as a product of all transitional probabilities between times $t=0$ and $t=(L+1)\tau$.
The most significant difference between the PDFs given by Eqs. (\ref{eq:Pv}) and (\ref{eq:Px}) is in the diffusion terms $C_v$, $C_x$ that appear in the denominator of the respective exponential factors. 
Specifically, after performing the Taylor expansion of $C_v$ and $C_x$ to the leading order, we obtain $C_v =  4 k_BT \Gamma\tau /m$ and $C_x =  \sfrac{8}{3}\, k_BT \Gamma \tau^3 /m$.
Here, the factor $\sfrac23$ that was discussed in \cite{FerrettiPRX20} appears directly.
 
\section{BEEPSIS inference from simulated trajectories}
\subsection{Harmonic oscillator}
\label{app:simho}
\begin{figure}
  \includegraphics[trim = {1.5cm 0 0 0},clip,width=\columnwidth]{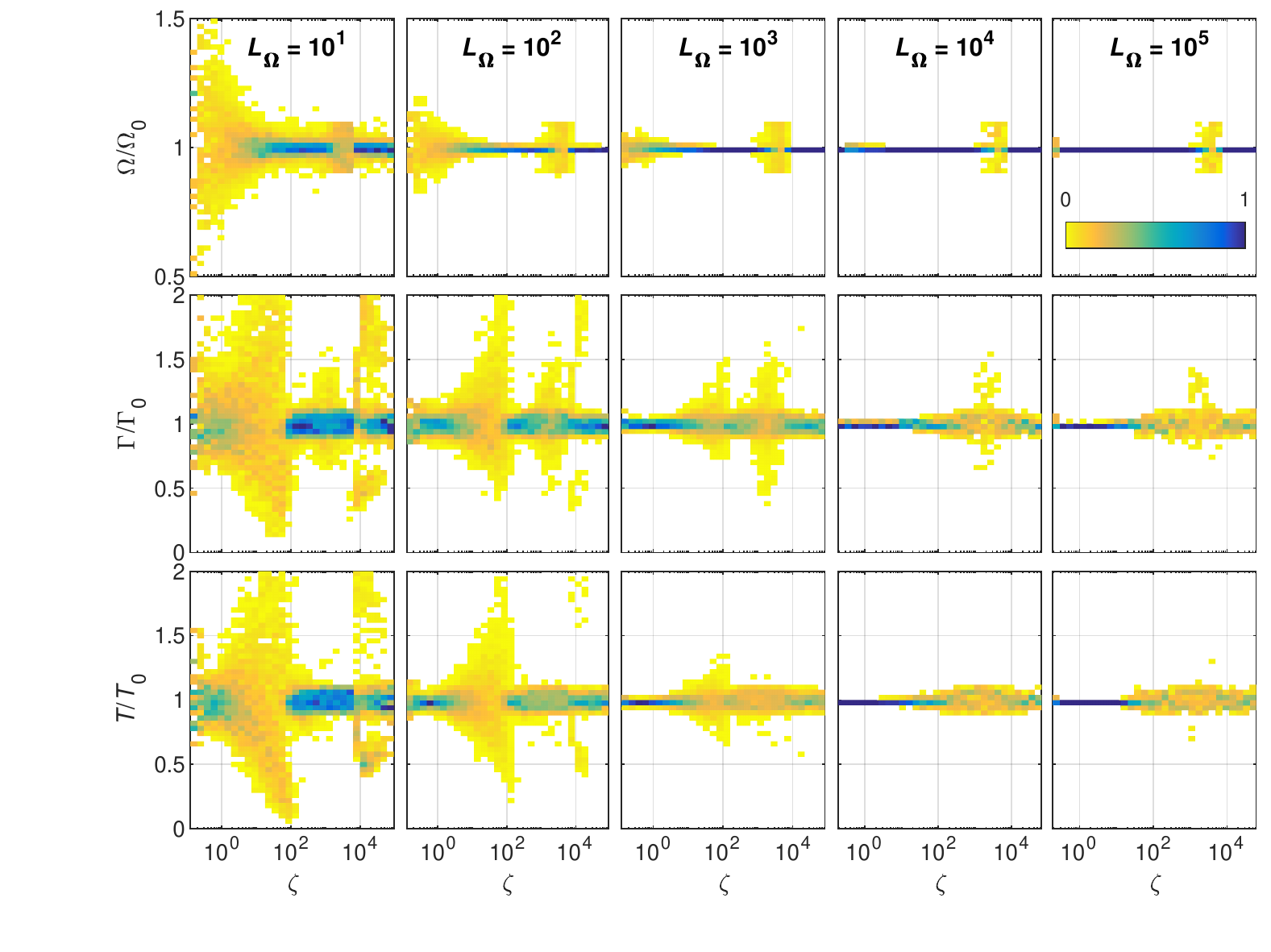}
  \caption{Performance of BEEPSIS on simulated harmonic oscillator trajectories as a function of the damping ratio of the oscillator $\zeta$. The same trajectories as used in generating Figs.~\ref{fig:HOsim}(d) - (f) were evaluated. From top to bottom, the rows correspond to the REI of $ \Omega / \Omega_0$, $ \Gamma / \Gamma_0 $, and $ T/T_0$. Each column then represents an ensemble of trajectories with a fixed length $L_\Omega$ (from left to right, $L_\Omega =$ 10, 100, 1000, 10$^4$ and 10$^5$ periods). For all analyzed trajectories, the sampling rate was fixed at $N_\Omega =  20$ points per oscillation period. 
  }
  \label{fig:HOreloud}
\end{figure}

In this Appendix, we extend the analysis of BEEPSIS performance on the simulated trajectories of harmonic oscillator, which was presented in Section \ref{sec:simul}. Here, we concentrate on the accuracy and precision of BEEPSIS inference in different regimes of particle motion ranging from overdamped to underdamped oscillations. 
The transition between these regimes of motion can be quantitatively described by the damping ratio $\zeta = 2\Omega_0/\Gamma_0$, with the values smaller than, equal to, and bigger than 1 representing an overdamped, critically damped, and underdamped oscillator, respectively. 
For each  of the simulated trajectories included in the analysis presented in Figs. \ref{fig:HOsim}(e) - (f) of the main text, we calculated the actual value of $\zeta$. Subsequently, we separated the corresponding ratios of estimated to input values (REIs) into 31 bins in $\zeta$, with equal width on the logarithmic scale.
Figure~\ref{fig:HOreloud} shows the distributions of the REIs of $ \Omega / \Omega_0$, $ \Gamma / \Gamma_0 $, and $ T/T_0$, evaluated as functions of $\zeta$ for five different lengths $L_\Omega$ of the simulated trajectories sampled with $N_\Omega = 20$ points per oscillation period. 
For each bin of $\zeta$, the spread of the distribution of each REI directly indicates the precision of inference of the particular parameter, whereas the accuracy of inference is reflected in the mean value of the corresponding distribution. 

One can see that for the underdamped cases with $\zeta>50$, the estimated oscillation eigenfrequency $\Omega$  is reasonably well determined even for trajectories as short as 10 periods. 
With increasing trajectory length, the precision of estimating $\Omega$ increases and its value can be correctly predicted even in the overdamped case with $\zeta<0.02$. 
We also observed an apparent instability in the determination of $\Omega$ for $\zeta$ in the range $\sim 10^3-10^4$. 
We were not able to determine its precise origin; however, it could likely be suppressed by a different choice of algorithm used for the minimization of the negative log likelihood in the BEEPSIS protocol or by fine tuning the minimization algorithm parameters (which had been set identical for all simulated trajectories). 
In contrast, the precision of the estimated damping rate $\Gamma$ and ambient temperature $T$ is rather low across all studied damping ratios for the shortest trajectory length of 10 periods. 
In the underdamped systems with stronger damping ($1< \zeta < 50$), the precision of estimation of $\Gamma$ and $T$ considerably improves for longer trajectories. 
However, even for the longest studied trajectories with $10^5$ periods, we can still see an increased spread in the REI of both $\Gamma$ and $T$ for underdamped systems with $\zeta > 100$. 
Intuitively, more deterministic behavior of such underdamped harmonic oscillators subject to random thermal forcing leads to a lower precision in inferring the diffusion-related environmental parameters $\Gamma$ and $T$ that are less strongly reflected in the system's dynamics.

\subsection{Duffing oscillator}
\label{app:simduff}
Similar to the case of harmonic oscillator (see Section \ref{sec:simharm} and Appendix \ref{app:simho}), we simulated the motion of a small particle (radius $a$ = 100 nm, density $\rho$ = 2000 $\mathrm{kg\, m^{-3}}$) confined in a trap with anharmonic Duffing force profile, $F_{\mathrm{D}}(x) = -m\Omega_0^2 x \left(1-\xi_0 x^2\right)$. 
The parameters $(\Omega_0, \xi_0, T_0)$ that enter into the corresponding Langevin equation were randomly selected from intervals $\Omega_0/(2\pi)\in\langle20,200\rangle$ kHz, $\xi_0 = \xi_0' \times 10^{-5}\Omega_0/(2\pi)$ where $  \xi_0'\in\langle 0.1, 5\rangle\, \mathrm{\mu m^{-2} s}$, 
and $T_0\in\langle10, 1000\rangle$ K. The viscous damping coefficient $\Gamma_0$ was then calculated according to~\cite{LiNatPhys11} for the given particle size and external air pressure randomly chosen in the range of $1-10^{5}$ Pa.
The trajectory of nonlinear Duffing oscillator might become unstable for some combinations of system parameters; therefore, for each simulated trajectory, we verified that the particle remained confined during the whole duration of the simulation. 
If that was not the case, a new set of simulation parameters was generated and a new trajectory was simulated. 
\begin{figure}[t]
  \includegraphics[width=\columnwidth]{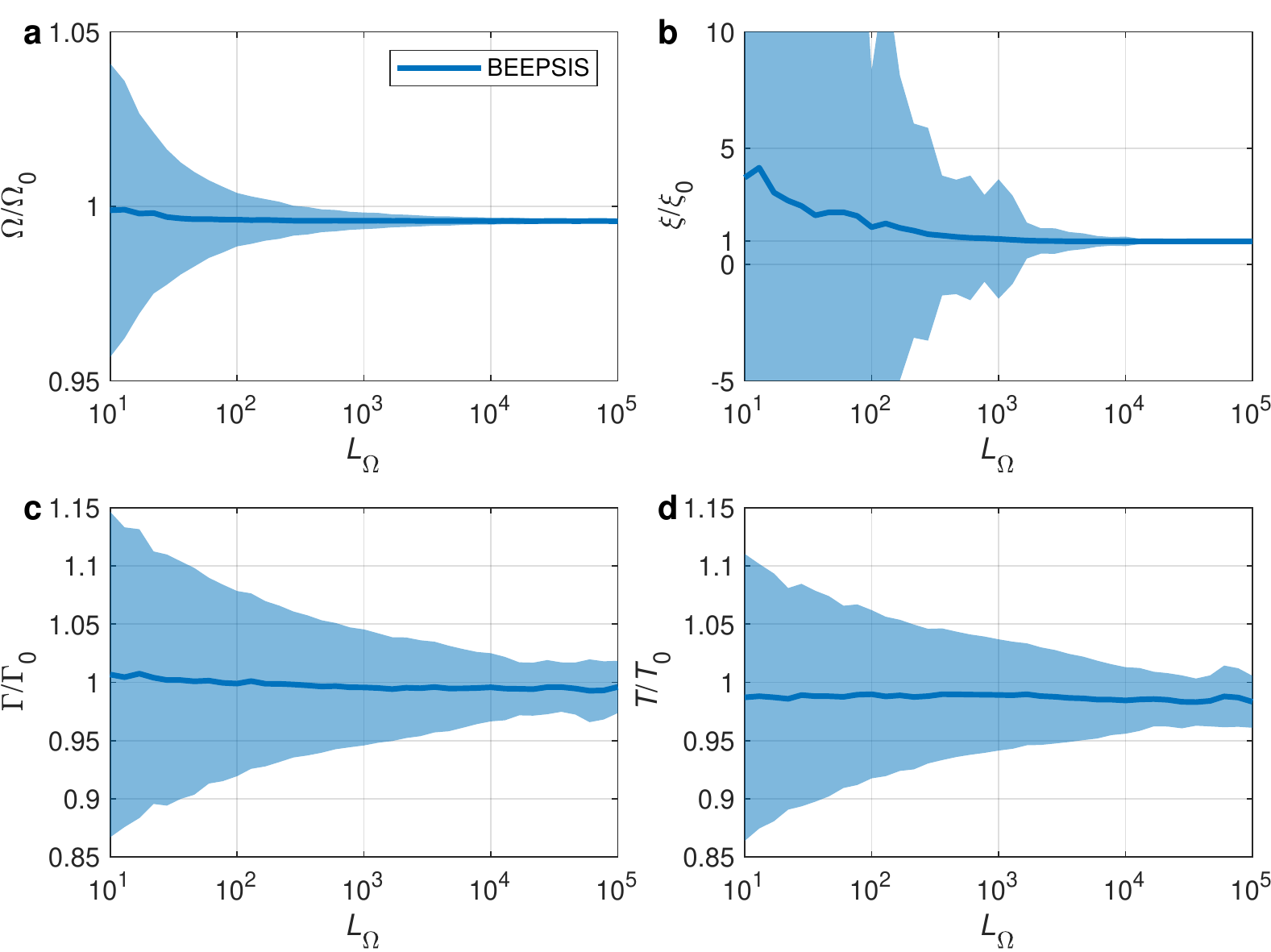}
  \caption{Performance of BEEPSIS in estimating the parameters of a stochastic Duffing oscillator from simulated trajectories. The plots show the accuracy (solid blue lines) and precision (blue shadings) of (a) harmonic eigenfrequency $\Omega$, (b) strength of Duffing nonlinearity $\xi$, (c) viscous damping rate $\Gamma$, and (d) temperature $T$ inferred by BEEPSIS as functions of the trajectory length $L_\Omega$ (see Section~\ref{sec:simharm} for the definition of the accuracy and precision of inference).
  For each trajectory length, $10^4$ trajectories were analyzed and the sampling rate was fixed at $N_\Omega =  20$ points per oscillation period for all trajectories.
 }
  \label{fig:DOsim}
\end{figure} 
\begin{figure}
  \includegraphics[trim = {1.5cm 0 0 0},clip,width=\columnwidth]{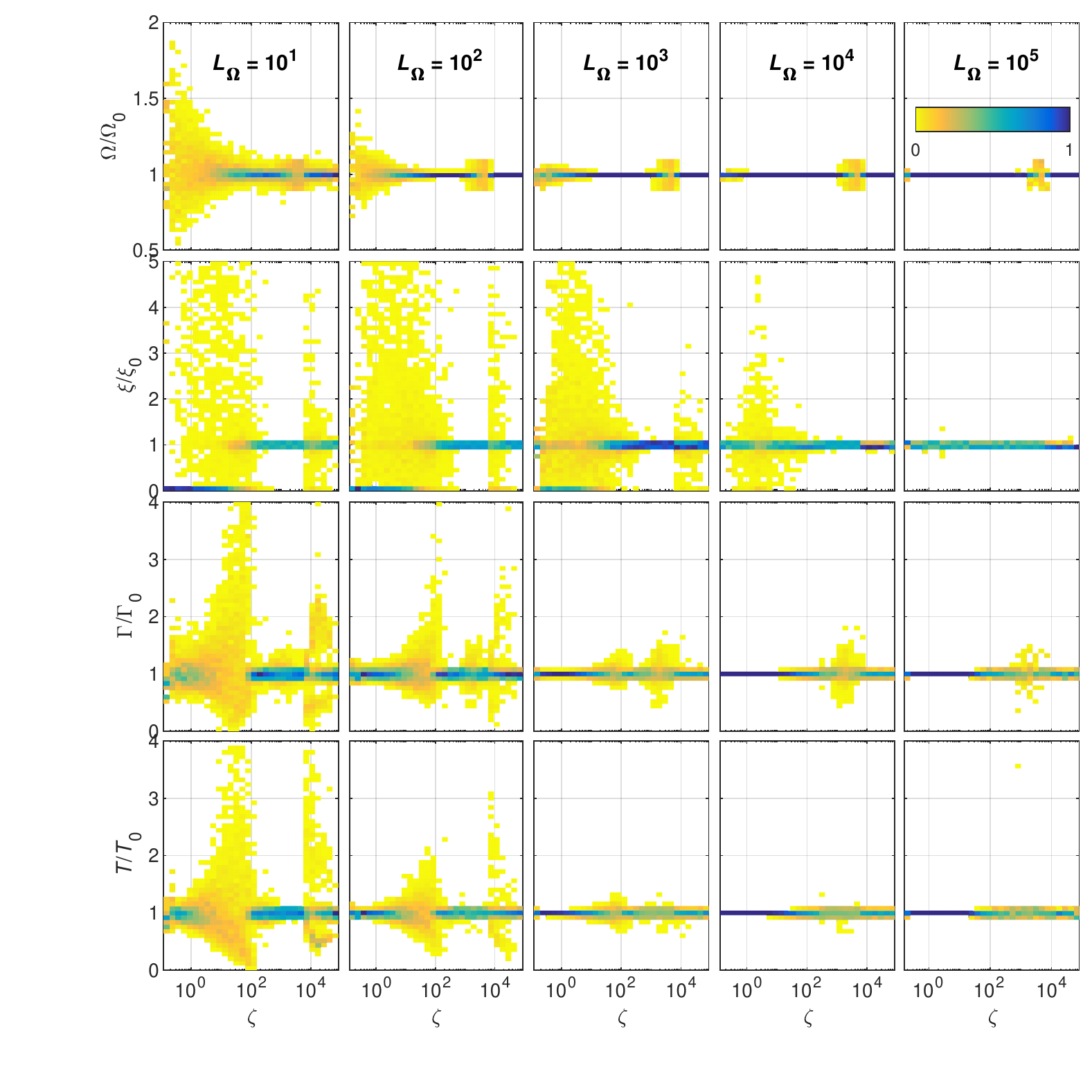}
  \caption{Performance of BEEPSIS on simulated Duffing oscillator trajectories as a function of the damping ratio of the oscillator $\zeta$. The same trajectories as used in generating Fig.~\ref{fig:DOsim} were evaluated. From top to bottom, the rows correspond to the REI of $ \Omega / \Omega_0$, $ \xi / \xi_0$, $ \Gamma / \Gamma_0 $, and $ T/T_0$. Each column then represents an ensemble of trajectories with a fixed length $L_\Omega$ (from left to right, $L_\Omega =$ 10, 100, 1000, 10$^4$ and 10$^5$ periods). For all analyzed trajectories, the sampling rate was fixed at $N_\Omega =  20$ points per oscillation period. 
  }
  \label{fig:DOreloud}
\end{figure}

Figure~\ref{fig:DOsim} shows the results of the BEEPSIS analysis of these trajectories presented in the same format as in Figs. \ref{fig:HOsim}(d) - (f), i.e., the accuracy (solid blue lines) and precision (blue shadings) of the inferred system parameters $(\Omega,\xi,\Gamma,T)$ are displayed as functions of the trajectory length $L_\Omega$. The trajectory sampling rate was kept constant at $N_\Omega = 20$ points per oscillation period in all simulations, with the oscillation period determined by the eigenfrequency $\Omega_0$ used in the simulations as $2\pi / \Omega_0$. 
As in the case of harmonic oscillator, BEEPSIS slightly underestimates the value of $\Omega$; this is reflected in the mean REI $\langle \Omega / \Omega_0 \rangle$, which quantifies the accuracy of estimation, being smaller than 1 by $\sim$\,0.2\%. 
The precision of estimation, given by the standard deviation of REIs inferred from the ensemble of trajectories simulated for the given value of $L_\Omega$, is then $<0.1$\,\% for trajectories with $L_\Omega > 8000$ periods.
The strength of Duffing nonlinearity $\xi$ is correctly determined for trajectories with $L_\Omega > 2000$ periods; specifically, for the longest studied trajectories with $L_\Omega = 10^5$ periods, $\xi$ is underestimated by $\sim$ 0.5\,\% with the precision of $\sim$ 6\,\%.
On the other hand, for shorter trajectories ($L_\Omega < 2000$ periods), the low precision of estimating $\xi$ precludes its reliable determination. 
The accuracy of estimation of $\Gamma$ and $T$ is rather good across the whole range of the studied trajectory lengths. Again, BEEPSIS slightly underestimates these parameters, with the bias $\sim$ 1\,\% for the viscous damping rate and $\sim1.5$\,\% for the temperature. 
In comparison with the harmonic oscillator case, the precision of estimation of these parameters is slightly lower, with the minimal standard deviations obtained for $L_\Omega = 10^5$ periods being $\sim$\,2.5\%.

Figure~\ref{fig:DOreloud} depicts the influence of the damping ratio $\zeta$ and the trajectory length $L_\Omega$ on the quality of BEEPSIS-inferred values of the Duffing oscillator parameters. 
Here, we can see that $\Omega$ and $\xi$, which characterize the confining force, are well determined for all studied trajectory lengths in underdamped systems with $\zeta > 100$, while in the overdamped regime, the precision of inference of $\Omega$ and especially of $\xi$ from short trajectories ($L_\Omega < 1000$ periods) significantly decreases.
Similar to the harmonic oscillator case, we also observed an instability in the determination of $\Omega$ for $\zeta$ in the range $\sim 10^3-10^4$. 
As discussed above, this instability could likely be reduced by specific tuning of the minimization algorithm for different values of $\zeta$.
Regarding the inference of $\Gamma$ and $T$, they could be determined with a good accuracy and precision from long trajectories ($L_\Omega > 100$ periods) recorded in the presence of appreciable damping ($\zeta<50$). However, in the highly underdamped regime with $\zeta>100$, the precision of inference decreases even for the longest studied trajectories with $10^5$ periods. 
This trend is similar to the behavior observed for harmonic oscillators (compare with Fig.~\ref{fig:HOreloud}); it is related to the smaller influence of environmental parameters on the dynamics of underdamped oscillators dominated by the inertial effects.

\section{Non-parametric estimation of force profile}
\label{app:binning}
In certain situations, especially when very little prior information is available about the spatial profile of the force acting on the studied stochastic system, it is useful to have a simple and fast method that gives a basic insight into the force profile. 
In devising such a simplified procedure for estimating the force, we will assume that the viscous damping rate is known (e.g., from a theoretical model or from a previous calibration measurement with a harmonic confinement force), the detection error is negligible, and the value of the effective ambient temperature is not of interest for the moment.  
Under these conditions, Eq.~(\ref{eq:bsum}) simplifies to 
\begin{equation}
 \mathcal{L} = \frac12 \sum_{n=x,y,\dots} \sum_{i,j=1}^{L}\bm{m}_{i,n} (\mathbf C^{-1})_{ij} \bm{m}_{j,n},
 \label{eq:bsum2}
\end{equation} 
with the misfits $\bm{m}_{i,j}$ and covariance matrix $\mathbf C$ defined by Eqs. (\ref{eq:misfit}--\ref{eq:Cb}).
In the next step, we divide the full extent of the particle's motion within the $N$-dimensional configuration space into $B$ discrete spatial bins and assume that the $N$-dimensional force vector is approximately constant in each of these bins, i.e., $\bm{F}_b = \mathrm{const.}$, where $b$ is the bin index. 
The assumption of stepwise-constant force profile leads to zero/undefined gradients of the force components. Thus, we cannot evaluate the influence of the detection error using Eqs. (\ref{eq:Pcerr}--\ref{eq:Pcerrend}) that explicitly depend on these force gradients.
After dividing the configuration space into bins, we find the corresponding bin index $b_i$ for each particle position $\bm{x}_i$ and define sets $S_b$ of all positions from the trajectory that fall into the given bin $b$. 
Using~(\ref{eq:misfit}), the misfits can be expressed as
\begin{equation}
  \bm{m}_i = \frac{\tau}{m\Gamma}\left(1-\mr e^{-\Gamma \tau}\right) \left(\bm{m}'_i  -  \bm{F}_{b_i}\right), 
\end{equation}
where
\begin{eqnarray}
\nonumber
  \bm{m'}_i &=& \left[\bm{x}_{i+1} -\bm{x}_i\left(1+ \mr e^{-\Gamma \tau}\right) + \bm{x}_{i-1}\mr e^{-\Gamma \tau}\right] \\
  &&\times \frac{m\Gamma}{\tau}\left(1-\mr e^{-\Gamma \tau}\right)^{-1}.
 \label{eq:misfit2}
\end{eqnarray} 
Upon differentiating~(\ref{eq:bsum2}) with respect to $\bm{F}_b$ and setting the result equal to zero, which is the condition required to minimize $\mathcal{L}$, we obtain a set of linear equations for $B$ unknown variables $\bm{F}_{b'}$
\begin{equation}
 \mathbf{Q}_{bb'} \bm{F}_{b'} = \bm{q}_b,
\end{equation} 
where
\begin{eqnarray}
 \label{eq:Q1}
 \mathbf{Q}_{bb'} &=& \sum\limits_{i\in S_b, j\in S_{b'}} \left(\mathbf C^{-1}\right)_{ij},  \\
 \bm{q}_b &=& \sum\limits_{i,j\in S_b} \left(\mathbf C^{-1}\right)_{ij} \bm{m'}_i.
 \label{eq:Q2}
\end{eqnarray}
In contrast to the tri-diagonal shape of $\mathbf{C}$, $\mathbf{C}^{-1}$ is, in principle, a full matrix whose elements can be expressed as~\cite{DaFonsecaLinAlgApp01}
\begin{equation}
\label{eq:Cinv1}
 (\mathbf{C}^{-1})_{ij} = (-1)^{i+j} \frac{r_{+}^{-|i-j|}}{r_{+}-r{-}} \times
    \left\{ \begin{array}{l}
              \mathcal{C}(i,j)\quad\mathrm{for\ } i\leq j\\
              \mathcal{C}(j,i)\quad\mathrm{for\ } i>j
             \end{array}
    \right. \\[1mm]
\end{equation}
where $r_{\pm} = a/(2b) \pm \sqrt{a^2/(4b^2) - 1}$ and
\begin{equation}
 \mathcal{C}(k,l) = \left[1-\left(\frac{r_{-}}{r_{+}}\right)^k\right] \left[1-\left(\frac{r_{-}}{r_{+}}\right)^{L-l+1}\right].
\end{equation}
Since the value of $r_{+}$ is positive and the exponent $-|i-j|$ of the first fraction numerator is always negative, the elements on the diagonals further away form the main diagonal quickly decrease and become smaller than the machine precision for $|i-j| \gtrsim 100$.
Moreover, the values of $\mathcal{C}$ become indistinguishable from 1 except for $i,j$ being placed in the top-left or bottom-right corners of matrix $\mathbf{C}^{-1}$.
This means that the elements on the diagonals are constant, with the exception of the values in the aforementioned corners. 
Therefore, one can easily evaluate the matrix $ \mathbf{Q}_{bb'} $ given by~(\ref{eq:Q1}) and vector $ \bm{q}_b$ given by (\ref{eq:Q2}).

A similar procedure can also be applied to the velocity-explicit PDFs defined by Eqs.~(\ref{eq:Pv}) and (\ref{eq:Px}). 
Due to the fact that the covariance matrix associated with these PDFs is purely diagonal, the forces can be directly expressed as 
\begin{eqnarray}
 \bm{F}_{v,b} &=& \frac{1}{N_b} \frac{m\Gamma}{ (1\!-\! \mr e^{-\Gamma \tau})}  \sum\limits_{i\in S_b} \left[ \bm{v}_{i+1}\! -\! \bm{v}_i \mr e^{-\Gamma \tau} \right],\\
 \nonumber
 \bm{F}_{x,b} &=& \frac{1}{N_b}  \frac{m\Gamma^2}{ (\Gamma \tau \!-\!1 \!+\!\mr e^{-\Gamma \tau} ) } \\
   && \times\sum\limits_{i\in S_b} \left[ \bm x_{\!i+\!1} \!-\! \bm{x}_i \! -\! \frac{\bm{v}_i}{\Gamma}\left(1\!-\!\mr e^{-\Gamma \tau}\right)\right],
\end{eqnarray}
where $N_b$ is the number of particle position occurrences in bin $b$.
The expressions for $\bm{F}_{v,b}$ and $\bm{F}_{x,b}$ are derived using the PDFs (\ref{eq:Pv}) and (\ref{eq:Px}), respectively, and are equivalent to each other assuming $\Gamma\tau\ll1$.
Estimation of the force profile from these quantities can be extremely fast and could possibly be used in ``real-time'' data processing schemes. However, before adopting this inference approach, its accuracy and precision for the given experimental scenario should first be quantified by simulations.

\section{Computer codes for trajectory processing}
We provide the MATLAB source codes for the BEEPSIS analysis of experimental trajectories on GITHUB, see \url{https://github.com/leviphot/BEEPSIS}.
The code requires licenses of MATLAB (release R2018a or newer) with Optimization Toolbox, Statistics Toolbox, and GADS Toolbox (optional). 
Time-critical part of the algorithm has been written in C++ language and requires compilation by the MATLAB \texttt{mex} compiler. 
Detailed instructions for code installation and description of the interface is provided in the file \texttt{readme.md} in the aforementioned loction.


\begin{thebibliography}{10}

\bibitem{ChandrasekharRMP43}
S.~Chandrasekhar.
\newblock {Stochastic Problems in Physics and Astronomy}.
\newblock {\em Rev. Mod. Phys.}, 15:1--89, 1943.

\bibitem{Wang_RMP_1945}
M.~C. Wang and G.~E. Uhlenbeck.
\newblock On the theory of the {B}rowian motion {II}.
\newblock {\em Rev. Mod. Phys.}, 17:323--342, 1945.

\bibitem{MoserNatNanotech13}
J.~Moser, J.~Guettinger, A.~Eichler, M.~J. Esplandiu, D.~E. Liu, M.~I. Dykman,
  and A.~Bachtold.
\newblock Ultrasensitive force detection with a nanotube mechanical resonator.
\newblock {\em Nature Nanotechnology}, 8(7):493--496, 2013.

\bibitem{RanjitPRA16}
Gambhir Ranjit, Mark Cunningham, Kirsten Casey, and Andrew~A. Geraci.
\newblock Zeptonewton force sensing with nanospheres in an optical lattice.
\newblock {\em {Phys. Rev. A}}, {93}({5}):{053801}, {2016}.

\bibitem{LassagneNanoLett08}
B.~Lassagne, D.~Garcia-Sanchez, A.~Aguasca, and A.~Bachtold.
\newblock Ultrasensitive mass sensing with a nanotube electromechanical
  resonator.
\newblock {\em Nano Letters}, 8(11):3735--3738, 2008.

\bibitem{TeufelNatNanotech09}
J.~D. Teufel, T.~Donner, M.~A. Castellanos-Beltran, J.~W. Harlow, and K.~W.
  Lehnert.
\newblock Nanomechanical motion measured with an imprecision below that at the
  standard quantum limit.
\newblock {\em Nature Nanotechnology}, 4(12):820--823, 2009.

\bibitem{TeufelNature2011}
J.~D. Teufel, T.~Donner, Dale Li, J.~W. Harlow, M.~S. Allman, K.~Cicak, A.~J.
  Sirois, J.~D. Whittaker, K.~W. Lehnert, and R.~W. Simmonds.
\newblock Sideband cooling of micromechanical motion to the quantum ground
  state.
\newblock {\em Nature}, 475(7356):359--363, 2011.

\bibitem{Tebbenjohanns_PRL_2020}
Felix Tebbenjohanns, Martin Frimmer, Vijay Jain, Dominik Windey, and Lukas
  Novotny.
\newblock Motional {Sideband} {Asymmetry} of a {Nanoparticle} {Optically}
  {Levitated} in {Free} {Space}.
\newblock {\em Phys. Rev. Lett.}, 124(1):013603, January 2020.

\bibitem{Millen_RPP_2020}
James Millen, Tania~S Monteiro, Robert Pettit, and A~Nick Vamivakas.
\newblock Optomechanics with levitated particles.
\newblock {\em Rep. Prog. Phys.}, 83(2):026401, jan 2020.

\bibitem{DelicScience20}
Uro{\v s} Deli{\'c}, Manuel Reisenbauer, Kahan Dare, David Grass, Vladan
  Vuleti{\'c}, Nikolai Kiesel, and Markus Aspelmeyer.
\newblock Cooling of a levitated nanoparticle to the motional quantum ground
  state.
\newblock {\em Science}, 367(6480):892--895, 2020.

\bibitem{Gieseler2012Subkelvin}
Jan Gieseler, Bradley Deutsch, Romain Quidant, and Lukas Novotny.
\newblock Subkelvin parametric feedback cooling of a laser-trapped
  nanoparticle.
\newblock {\em Phys. Rev. Lett.}, 109(10), 2012.

\bibitem{TebbenjohannsPRL19}
F.~Tebbenjohanns, M.~Frimmer, A.~Militaru, V.~Jain, and L.~Novotny.
\newblock Cold damping of an optically levitated nanoparticle to microkelvin
  temperatures.
\newblock {\em Phys. Rev. Lett.}, 122:223601--6, 2019.

\bibitem{CurtisOPTCOMM02}
J.~E. Curtis, B.~A. Koss, and D.~G. Grier.
\newblock {Dynamic holographic optical tweezers}.
\newblock {\em Opt. Commun.}, 207:169--175, 2002.

\bibitem{CizmarLPL11}
T.~\v{C}i\v{z}m{\'a}r, O.~Brzobohat{\'y}, K.~Dholakia, and P.~Zem{\'a}nek.
\newblock {The holographic optical micro-manipulation system based on
  counter-propagating beams}.
\newblock {\em Laser Phys. Lett.}, 8(1):50--56, 2011.

\bibitem{SilerPRL18}
M.~\v{S}iler, L.~Ornigotti, O.~Brzobohat\'{y}, P.~J\'{a}kl, A.~Ryabov,
  V.~Holubec, P.~Zem\'{a}nek, and R.~Filip.
\newblock Diffusing up the hill: Dynamics and equipartition in highly unstable
  systems.
\newblock {\em Phys. Rev. Lett.}, 121:23601, 2018.

\bibitem{SiegertPLA1998}
S~Siegert, R~Friedrich, and J~Peinke.
\newblock Analysis of data sets of stochastic systems.
\newblock {\em Physics Letters A}, 243(5-6):275--280, 1998.

\bibitem{RagwitzPRL2001}
M~Ragwitz and H~Kantz.
\newblock Indispensable finite time corrections for fokker-planck equations
  from time series data.
\newblock {\em Physical Review Letters}, 87(25), 2001.

\bibitem{Jonesbook15}
P.~Jones, O.~Marag{\`o}, and G.~Volpe.
\newblock {\em {Optical tweezers: Principles and Applications}}.
\newblock Cambridge University Press, Cambridge, 2015.

\bibitem{GieselerAOP2021}
Jan Gieseler, Juan~Ruben Gomez-Solano, Alessandro Magazzu, Isaac~Perez
  Castillo, Laura~Perez Garcia, Marta Gironella-Torrent, Xavier Viader-Godoy,
  Felix Ritort, Giuseppe Pesce, Alejandro Arzola, V, Karen Volke-Sepulveda, and
  Giovanni Volpe.
\newblock Optical tweezers - from calibration to applications: a tutorial.
\newblock {\em Advances in Optics and Photonics}, 13(1):74--241, 2021.

\bibitem{FlorinAPA98}
E.-L. Florin, A.~Pralle, E.~H.~K. Stelzer, and J.~K.~H. H{\"o}rber.
\newblock {Photonic force microscope calibration by thermal noise analysis.}
\newblock {\em Appl. Phys. A}, 66:75--78, 1998.

\bibitem{BergSorensenRSI04}
K.~Berg-S{\o}rensen and H.~Flyvbjerg.
\newblock {Power spectrum analysis for optical tweezers}.
\newblock {\em Rev. Sci. Instrum.}, 75:594--612, 2004.

\bibitem{RichlyOE13}
Maximilian~U. Richly, Silvan T\"{u}rkcan, Antoine~Le Gall, Nicolas Fiszman,
  Jean-Baptiste Masson, Nathalie Westbrook, Karen Perronet, and Antigoni
  Alexandrou.
\newblock Calibrating optical tweezers with bayesian inference.
\newblock {\em Opt. Express}, 21(25):31578--31590, 2013.

\bibitem{PerezGarciaNatComm18}
Laura Perez~Garcia, Jaime Donlucas~Perez, Giorgio Volpe, Alejandro~V. Arzola,
  and Giovanni Volpe.
\newblock High-performance reconstruction of microscopic force fields from
  brownian trajectories.
\newblock {\em Nat.Commun.}, 9, 2018.

\bibitem{FrishmanPRX20}
Anna Frishman and Pierre Ronceray.
\newblock Learning force fields from stochastic trajectories.
\newblock {\em Phys. Rev. X}, 10:021009, 2020.

\bibitem{Strogatz}
S.H Strogatz.
\newblock {\em Nonlinear Dynamics and Chaos with Applications to Physics,
  Biology, Chemistry and Engineering}.
\newblock Westview Press, Boulder, 2015.

\bibitem{LehlePRE2015}
B.~Lehle and J.~Peinke.
\newblock Analyzing a stochastic time series obeying a second-order
  differential equation.
\newblock {\em Physical Review E}, 91(6), 2015.

\bibitem{FerrettiPRX20}
Federica Ferretti, Victor Chard\`es, Thierry Mora, Aleksandra~M. Walczak, and
  Irene Giardina.
\newblock Building general langevin models from discrete datasets.
\newblock {\em Phys. Rev. X}, 10:031018, Jul 2020.

\bibitem{BrucknerPRL20}
David~B. Br\"uckner, Pierre Ronceray, and Chase~P. Broedersz.
\newblock Inferring the dynamics of underdamped stochastic systems.
\newblock {\em Phys. Rev. Lett.}, 125:058103, 2020.

\bibitem{FlajsmanovaSR20}
Jana Flaj\v{s}manov\'{a}, Martin \v{S}iler, Petr Jedli\v{c}ka, Franti\v{s}ek
  Hrub\'{y}, Oto Brzobohat\'{y}, Radim Filip, and Pavel Zem\'{a}nek.
\newblock Using the transient trajectories of an optically levitated
  nanoparticle to characterize a stochastic {D}uffing oscillator.
\newblock {\em Scientific Reports}, 10:14436, 2020.

\bibitem{SvakNC18}
V.~Svak, O~Brzobohat\'{y}, M~\v{S}iler, P~J\'{a}kl, J~Ka\v{n}ka, P~Zem\'{a}nek,
  and SH~Simpson.
\newblock {Transverse spin forces and non-equilibrium particle dynamics in a
  circularly polarized vacuum optical trap}.
\newblock {\em {Nat. Commun.}}, {9}:{5453}, {2018}.

\bibitem{Risken}
H.~Risken.
\newblock {\em {The Fokker-Planck Equation}}.
\newblock Springer-Verlag, Berlin, 1996.

\bibitem{SvakOptica21}
Vojt\v{e}ch Svak, Jana Flaj{\v s}manov{\'a}, Luk{\'a}{\v s} Chv{\'a}tal, Martin
  {\v S}iler, Alexandr Jon{\'a}{\v s}, Jan Je{\v z}ek, Stephen~H. Simpson,
  Pavel Zem{\'a}nek, and Oto Brzobohat{\'y}.
\newblock Stochastic dynamics of optically bound matter levitated in vacuum.
\newblock {\em Optica}, 8:220{\textendash}229, Feb 2021.

\bibitem{RieserScience22}
Jakob Rieser, Mario~A. Ciampini, Henning Rudolph, Nikolai Kiesel, Klaus
  Hornberger, Benjamin~A. Stickler, Markus Aspelmeyer, and Uro\v{s} Deli\'c.
\newblock Tunable light-induced dipole-dipole interaction between optically
  levitated nanoparticles.
\newblock {\em Science}, 377(6609):987--990, 2022.

\bibitem{DoiPolymerBook}
M.~Doi and S.~F. Edwards.
\newblock {\em {The Theory of Polymer Dynamics}}.
\newblock Oxford University Press, Oxford, 1986.

\bibitem{ErmakJCompPhys80}
Donald~L. Ermak and Helen Buckholz.
\newblock Numerical integration of the langevin equation: Monte carlo
  simulation.
\newblock {\em Journal of Computational Physics}, 35(2):169--182, 1980.

\bibitem{GronbechJensenMolPhys13}
Niels Gr{\o}nbech-Jensen and Oded Farago.
\newblock A simple and effective verlet-type algorithm for simulating langevin
  dynamics.
\newblock {\em Molec. Phys.}, 111(8):983--991, 2013.

\bibitem{SiviaBook}
D.~S. Sivia and J.~Skilling.
\newblock {\em {Data Analysis - A Bayesian Tutorial}}.
\newblock Oxford Science Publications. Oxford University Press, 2nd edition,
  2006.

\bibitem{TurkcanBJ12}
Silvan T{\"u}rkcan, Jean-Baptiste Masson, Didier Casanova, Genevi{\`e}ve
  Mialon, Thierry Gacoin, Jean-Pierre Boilot, Michel~R. Popoff, and Antigoni
  Alexandrou.
\newblock Observing the confinement potential of bacterial pore-forming toxin
  receptors inside rafts with nonblinking Eu3+-doped oxide nanoparticles.
\newblock {\em Biophys. J.}, 102(10):2299--2308, 2012.

\bibitem{GieselerNatPhys13}
Jan Gieseler, Lukas Novotny, and Romain Quidant.
\newblock {Thermal nonlinearities in a nanomechanical oscillator}.
\newblock {\em Nat. Phys.}, 9:{806--810}, 2013.

\bibitem{RevModPhysHanggi2009}
Peter H\"anggi and Fabio Marchesoni.
\newblock Artificial brownian motors: Controlling transport on the nanoscale.
\newblock {\em Rev. Mod. Phys.}, 81:387--442, 2009.

\bibitem{ToeplitzDet}
user17762.
\newblock How to compute the determinant of a tridiagonal matrix with constant
  diagonals?
\newblock Mathematics Stack Exchange.
\newblock URL:https://math.stackexchange.com/q/267466 (version: 2015-02-15).

\bibitem{DaFonsecaLinAlgApp01}
C.M. {da Fonseca} and J.~Petronilho.
\newblock Explicit inverses of some tridiagonal matrices.
\newblock {\em Linear Algebra and its Applications}, 325(1):7--21, 2001.

\bibitem{NorrelykkePRE11}
Simon~F. N\o{}rrelykke and Henrik Flyvbjerg.
\newblock Harmonic oscillator in heat bath: Exact simulation of
  time-lapse-recorded data and exact analytical benchmark statistics.
\newblock {\em Phys. Rev. E}, 83:041103, 2011.

\bibitem{LiNatPhys11}
Tongcang Li, Simon Kheifets, and Mark~G. Raizen.
\newblock {Millikelvin cooling of an optically trapped microsphere in vacuum}.
\newblock {\em Nat. Physics}, 7(7):527--530, 2011.

\bibitem{Gieseler_entropy_2018}
Jan Gieseler and James Millen.
\newblock Levitated {Nanoparticles} for {Microscopic} {Thermodynamics}?{A}
  {Review}.
\newblock {\em Entropy}, 20(5):326, 2018.

\bibitem{KramersPHYS40}
H.~A. Kramers.
\newblock {Brownian motion in the field of force and the diffusion model of
  chemical reactions}.
\newblock {\em Physica}, 7(4):284--304, 1940.

\bibitem{ReimannPhysRep02}
Peter Reimann.
\newblock {Brownian Motors: noisy transport far from equilibrium}.
\newblock {\em Phys. Rep.}, 361:57--265, 2002.

\bibitem{Rondin2017}
Lo{\"i}c Rondin, Jan Gieseler, Francesco Ricci, Romain Quidant, Christoph
  Dellago, and Lukas Novotny.
\newblock Direct measurement of kramers turnover with a levitated nanoparticle.
\newblock {\em Nat. Nanotechnol.}, 12:1130, 2017.

\bibitem{NeumeierArxiv22}
Lukas Neumeier, Mario~A. Ciampini, Oriol Romero-Isart, Markus Aspelmeyer, and
  Nikolai Kiesel.
\newblock Fast quantum interference of a nanoparticle via optical potential
  control, 2022.

\bibitem{MagriniNAT21}
Lorenzo Magrini, Philipp Rosenzweig, Constanze Bach, Andreas Deutschmann-Olek,
  Sebastian~G. Hofer, Sungkun Hong, Nikolai Kiesel, Andreas Kugi, and Markus
  Aspelmeyer.
\newblock Real-time optimal quantum control of mechanical motion at room
  temperature.
\newblock {\em Nature}, 595(7867):373+, 2021.

\bibitem{TebbenjohannsNAT21}
Felix Tebbenjohanns, M.~Luisa Mattana, Massimiliano Rossi, Martin Frimmer, and
  Lukas Novotny.
\newblock Quantum control of a nanoparticle optically levitated in cryogenic
  free space.
\newblock {\em Nature}, 595(7867):378+, 2021.

\bibitem{YonedaJPB17}
M.~Yoneda and K.~Aikawa.
\newblock Thermal broadening of the power spectra of laser-trapped particles in
  vacuum.
\newblock {\em J. Phys. B: At. Mol. Opt. Phys.}, 50:245501--9, 2017.

\bibitem{HebestreitRSI18}
Erik Hebestreit, Martin Frimmer, Ren\'{e} Reimann, Christoph Dellago, Francesco
  Ricci, and Lukas Novotny.
\newblock Calibration and energy measurement of optically levitated
  nanoparticle sensors.
\newblock {\em Rev. Sci. Instrum.}, 89(3):033111, 2018.

\bibitem{Grier2003}
David~G Grier.
\newblock A revolution in optical manipulation.
\newblock {\em Nature}, 424(6950):810, 2003.

\bibitem{GammaitoniRMP98}
Luca Gammaitoni, Peter H{\"a}nggi, Peter Jung, and Fabio Marchesoni.
\newblock {Stochastic resonance}.
\newblock {\em Rev. Mod. Phys.}, 70:223--287, 1998.

\bibitem{Soper2008Classical}
DE~Soper.
\newblock {\em Classical Field Theory}.
\newblock Dover Publications Inc., 2008.

\bibitem{Antognozzi2016Direct}
M.~Antognozzi, C.~R. Bermingham, R.~L. Harniman, S.~Simpson, J.~Senior,
  R.~Hayward, H.~Hoerber, M.~R. Dennis, A.~Y. Bekshaev, K.~Y. Bliokh, and
  F.~Nori.
\newblock Direct measurements of the extraordinary optical momentum and
  transverse spin-dependent force using a nano-cantilever.
\newblock {\em Nat. Phys.}, 12(8):731--735, 2016.

\bibitem{BhatiaIEEE13}
Harsh Bhatia, Gregory Norgard, Valerio Pascucci, and Peer-Timo Bremer.
\newblock The {H}elmholtz-{H}odge decomposition -- a survey.
\newblock {\em IEEE Transactions on Visualization and Computer Graphics},
  19(8):1386--1404, 2013.

\bibitem{NorrelykkeRSI2010}
S.~F. N{\o}rrelykke and H.~Flyvbjerg.
\newblock Power spectrum analysis with least-squares fitting: Amplitude bias
  and its elimination, with application to optical tweezers and atomic force
  microscope cantilevers.
\newblock {\em Review of Scientific Instruments}, 81, 2010.

\bibitem{OksendalBook}
Bernt Oksendal.
\newblock {\em Stochastic Differential Equations (3rd Ed.): An Introduction
  with Applications}.
\newblock Springer-Verlag, Berlin, Heidelberg, 2003.

\end{thebibliography}

\end{document}